\documentclass[twocolumn,showpacs,preprintnumbers,amsmath,amssymb,superscriptaddress]{revtex4}
\usepackage{textcomp}
\usepackage[pdftex]{epsfig}
\DeclareGraphicsExtensions{.jpg,.mps,.pdf,.png,.eps}

\usepackage{hyperref}

\usepackage{dcolumn}
\usepackage{bm}
\usepackage{comment}
\usepackage{color}

\usepackage[pdftex]{epsfig}
\DeclareGraphicsExtensions{.jpg,.mps,.pdf,.png}

\newcommand{\mA}{{\cal A}}
\newcommand{\mB}{{\cal B}}

\newcommand{\mD}{{\cal D}}
\newcommand{\mP}{{\cal P}}

\newcommand{\vx}{\textbf{x}}
\newcommand{\vk}{\textbf{k}}

\newcommand{\dd}{\hbox{d}}
\newcommand{\ii}{\hbox{i}}

\newcommand{\hXi}{{\hat{\Xi}}}
\newcommand{\hvarphi}{{\hat{\varphi}}}
\newcommand{\htau}{{\hat{\tau}}}

\newcommand{\mgen}{{\cal M}}
\newcommand{\hrho}{{\hat \rho}}
\newcommand{\hs}{{\hat{s}}}
\newcommand{\Dr}{{\Delta R}}

\newcommand{\eff}[1]{#1_{\rm eff}}
\newcommand{\init}{{\rm init}}
\newcommand{\tW}{{\tilde W}}
\newcommand{\lin}{{\rm lin}}
\newcommand{\cell}{{\rm cell}}

\usepackage{color}
\definecolor{Red}{rgb}{0.65,0.08,0.05}
\definecolor{Blue}{rgb}{0.05,0.08,0.65}
\definecolor{Purple}{RGB}{143,51,143}

\begin{document}
\title{Statistics  of cosmic density profiles from perturbation theory }
\author{Francis Bernardeau}
\affiliation{Institut de Physique Th\'eorique, CEA, IPhT, F-91191 Gif-sur-Yvette,\\
CNRS, URA 2306, F-91191 Gif-sur-Yvette, France}
\author{Christophe Pichon}
\affiliation{Institut d'Astrophysique de Paris \& UPMC (UMR 7095), 98 bis boulevard Arago, 75014 Paris, France}
\affiliation{Institute of Astronomy, University of Cambridge, Madingley Road, Cambridge, CB3 0HA, United Kingdom}
\author{Sandrine Codis}
\affiliation{Institut d'Astrophysique de Paris \& UPMC (UMR 7095), 98 bis boulevard Arago, 75014 Paris, France}
\begin{abstract}
{
The joint probability distribution function (PDF) of the density within multiple concentric spherical cells is considered. It is shown how its cumulant generating function can be obtained  at tree order in perturbation theory as the Legendre transform of a function directly built in terms of the initial moments. In the context of the upcoming  generation of  large-scale structure surveys, it is  conjectured that this result correctly models such a function for finite values of the variance. Detailed consequences of this assumption are explored. In particular the corresponding one-cell density probability distribution at finite variance is  computed for realistic power spectra, taking into account its  scale variation. It is found to be in agreement with $\Lambda$-CDM simulations at the few percent level for a wide range of density values and parameters.  Related explicit analytic expansions at the low and high density tails are given. The conditional (at fixed density) and marginal probability of the slope -- the density difference between adjacent cells -- and its fluctuations is also computed from the two-cells joint PDF; it also compares very well to simulations, in particular in under-dense regions, with a significant reduced cosmic scatter compared  to over-dense regions. It is emphasized that this could prove useful when studying the statistical properties of voids as it can serve as a statistical indicator to test gravity models and/or probe key cosmological parameters. 
}
\end{abstract}
\vskip2pc

\maketitle
\section{Introduction}

With new generations of surveys  either from ground-based facilities (eg BigBOSS,  DES,   Pan-STARRS, 
LSST~\footnote{\texttt{http://bigboss.lbl.gov},\, \\
\texttt{https://www.darkenergysurvey.org},\,
\texttt{http://pan-starrs.ifa.hawaii.edu},\,
\texttt{http://www.lsst.org}}) 
or space-based observatories (EUCLID~\citep{Euclid}, SNAP and JDEM~\footnote{\texttt{http://sci.esa.int/euclid},\\
 \texttt{http://snap.lbl.gov},\,\\ \texttt{http://jdem.lbl.gov}}), 
it will be possible to test with unprecedented accuracy the details of gravitational instabilities, in particular 
as it enters the nonlinear regime. These confrontations can be used in principle to test the gravity models (see for instance \cite{2010PhRvD..81l3508D,2007MNRAS.380.1029H}) and/or more generally improve upon our 
knowledge of cosmological parameters as detailed in \citep{Euclid}.

There are  only a limited range of quantities that can be computed from first principles. Next-to-leading order terms to power spectra and poly-spectra have been investigated extensively over the last few years with the introduction of novel methods.
Standard perturbation theory (PT) calculations, as described in \cite{2002PhR...367....1B}, have indeed 
been extended by the development of alternative analytical methods that 
try to improve upon standard  calculations. The first significant progress in this line of calculations in the Renormalized Perturbation Theory (RPT) 
proposition  \cite{2006PhRvD..73f3519C} followed by the closure theory \cite{2008ApJ...674..617T} and the time flow equations approach 
\cite{2008JCAP...10..036P}. Latest propositions, namely MPTbreeze \cite{2012MNRAS.427.2537C} and RegPT \cite{2012PhRvD..86j3528T} 
incorporate 2-loop order calculations and are accompanied by publicly released codes.  Recent developments involve the
effective field theory approaches \cite{2012arXiv1206.2926C}.

Alternatively one may look for  more global properties of the fields that capture some aspects of their non-Gaussian nature. A number of tests
have been put forward from peak statistics, see \cite{1986ApJ...304...15B}, that set the stage for Gaussian fields,  to topological invariants. The latter, introduced for instance in  \cite{1988PASP..100.1307G} or in \cite{1994A&A...288..697M} aim at 
producing robust statistical indicators. This topic was renewed in  \cite{2000PhRvL..85.5515C} and \cite{2012PhRvD..85b3011G}
with  the introduction of the notion of skeleton.
How such observables are affected by weak deviations from
Gaussianity was investigated originally in \cite{1994ApJ...434L..43M}  and for instance more recently in \citep{2012PhRvD..85b3011G,Codis2013} with the use of standard tools such as the Edgeworth expansion applied here to multiple variable distributions. These approaches, although promising, 
are hampered by the limited range of applicability of such expansions and as a consequence have to be  restricted to a limited
range of parameters and are usually confined  to the non-rare event region.

There is however at least one counter-example to that general statement:  the density probability distribution functions in concentric cells. As we will show in details in the following it is possible to get a global picture of what the joint density PDF should be, including in its rare event tails. 
 The size of the past surveys prevented an effective use of such statistical tools. Their current size
 makes it now possible to try and confront theoretical calculations with observations. 

Hence the aim of the paper  to revisit these calculations and assess their  domain
of validity with the help of numerical simulations.

To a large extent, the mathematical foundations of the calculation of the 
density probability distribution functions in concentric cells are to be  found in early works by
  Balian and Schaeffer \cite{1989A&A...220....1B}, who explored the connexion between count-in-cells statistics and the properties
 of the cumulant generating functions. In that  paper, the shape of the latter was just assumed without direct connexion with the dynamical equations. 
 This connexion was established in \cite{1992ApJ...392....1B} where it was shown that the leading order generating function of the count-in-cells probability distribution function could be derived from the dynamical equations. More precise calculations  were 
 developed in a systematic way in \cite{1994A&A...291..697B}, that take into account filtering effects, as pioneered in \cite{1993ApJ...412L...9J,1994ApJ...433....1B} where the impact of a Gaussian 
 window function or a top-hat window function was taken into account. At the same time, these predictions
were confronted to simulations and shown to be in excellent agreement with the numerical results (see for instance \cite{1994A&A...291..697B,1995MNRAS.274.1049B}). We will revisit here the quality of these predictions with the help of more 
accurate simulations. 
In parallel, it was shown that the same formalism could address more varied situations: large-scale biasing in \cite{1996A&A...312...11B}, projection effects in \cite{1995A&A...301..309B,2000A&A...364....1B}. A comprehensive presentation of these early works can be found in \cite{2002PhR...367....1B}.

Insights into the theoretical foundations of this approach were presented in \cite{2002A&A...382..412V} that
 allow to go beyond the diagrammatic approach
that was initially employed. The key argument is that for densities in concentric cells, {\sl the leading contributions}
in the implementation of the steepest descent method to the integration over field configurations should be 
configurations {\sl that are spherically symmetric}.
One can then take advantage of  Gauss' theorem to map the final field configuration into the initial one with a finite number of initial variables, on a cell-by-cell basis.
This is the strategy we  adopt below.  The purpose of this work is  to re-derive the fundamental relation that was obtained by the above mentioned authors, and to revisit the practical implementation of these calculations alleviating some of the shortcuts that were used in the literature. 

Specifically, the first objective of this paper is to quantity the sensitivity of the predictions for the one-cell PDF for the density on the power spectrum shape, its index and the scale-dependence of the latter (the so-called running parameter). 
The second objective  is to show that it is possible to use the  two-cells formalism to derive the statistical properties of the density {\em slope} defined as the difference of the density in two concentric cells of
 (possibly infinitesimally) close radii and more globally the whole density profile.
More specifically we show that for sufficiently steep power spectra (index less than $-1$), it is possible to take the limit 
of infinitely close top-hat radii and define the density slope at a given radius. We can then take advantage of this machinery to derive 
low-order cumulants of this quantity as well as its complete PDF. 
Finally this investigation allows us to  make a theoretical connexion with recent efforts (see for instance 
\cite{2012ApJ...754..109L,Sutter2012,Bolejko2011,van-de-Weygaert2011,Aragon-Calvo2010,Lavaux2010,2013MNRAS.428.3409A}) in exploring the low-density  regions
and their properties~\footnote{specifically, we  expect perturbation theory to break down later in low density regions; we also expect 
these regions to probe smaller Lagrangian scales.} such as the constrained average slope and its fluctuations given the (possibly low) value of the local density. This opens a 
way to exploit the properties of low-density regions: we will suggest that the expected profile of low-density regions is in fact a robust tool to use when matching theoretical predictions to  catalogues.

The outline of the paper is the following. In Section II we present the general formalism of how the cumulant generating functions are related to 
the spherical collapse dynamics. In Section III, this relationship is applied to derive the one-point density PDF;  the sensitivity of the predictions with scale and with the
power spectrum shape is  also reviewed there. In section IV,  we define the density profile and the slope, and derive its statistical properties. 
A summary and discussion on the scope of these results is given in the last section.

\section{The cumulant generating function at tree order}

Let us first revisit the derivation of the tree-order cumulant generating functions for densities computed in concentric cells.

\subsection{Definitions and connexions to   spherical collapse}
We consider a cosmological density field, $\rho(\vx)$, which is statically isotropic and homogeneous. The average value of $\rho(\vx)$ 
is set to unity. We then consider a random position $\vx_{0}$ and $n$ concentric cells of radius $R_{i}$ centered on $\vx_{0}$. 
The densities, $\rho_{i}$, obtained as the density within the radius $R_{i}$,
\begin{equation}
\rho_{i}=\frac{1}{4\pi R_{i}^{3}/3}\int_{\vert\vx-\vx_{0}\vert<R_{i}}\dd^{3}\vx\ \rho(\vx),
\end{equation}
form a set of correlated random variables. For a non-linearly evolved cosmic density field, they display non-Gaussian statistical properties. 
It is therefore natural to define the generating function of their joint moments as
\begin{equation}
\mgen(\{\lambda_{k}\})=\sum_{p_{i}=0}^{\infty}\ \langle \Pi_{i}\rho_{i}^{p_{i}}\rangle \frac{\Pi_{i}\lambda_{i}^{p_{i}}}{\Pi_{i}p_{i}!}\,,
\label{mdef}
\end{equation}
which can be simply expressed as
\begin{equation}
\mgen(\{\lambda_{k}\})=\langle \exp(\sum_{i}\lambda_{i}\rho_{i})\rangle.
\label{mdef2}
\end{equation}
The generating function, $\mgen(\{\lambda_{k}\})$, is a function of the $n$ variables $\lambda_{k}$. 
A very general theorem (see for instance \cite{BDFN92,Heyde63}) states that
this generating function is closely related to the joint \textsl{cumulant} generating function,
\begin{equation}
\varphi(\{\lambda_{k}\})=\sum_{p_{i}=0}^{\infty}\ \langle \Pi_{i}\rho_{i}^{p_{i}}\rangle_{c}\frac{\Pi_{i}\lambda_{i}^{p_{i}}}{\Pi_{i}p_{i}!}\,,
\label{phidef}
\end{equation}
via the relation
\begin{equation}
\mgen(\{\lambda_{k}\})=\exp\left[ \varphi(\{\lambda_{k}\})\right]\,.
\end{equation}
Note importantly that this makes $\varphi(\{\lambda_{k}\})$ an observable on its own \footnote{It is to be noted however that  it is not necessarily defined for any real values 
of $\lambda$ as one expects for instance that the 1-point density PDF develops an exponential cut-off, $\mP(\rho)\sim \exp(-\lambda_{c}\rho)$, which implies that $\varphi(\lambda)$ is undefined when $\lambda>\lambda_{c}$.}.

The tree order expression of such cumulants can be derived from a direct expansion of the density field, i.e.
\begin{equation}
\rho(\vx)=1+\delta^{(1)}+\delta^{(2)}+\delta^{(3)}+\dots\,,
\end{equation}
where $\delta^{(p)}$ is of order $p$ with respect to the initial density contrast. For Gaussian initial conditions 
the leading order cumulant (that is the connected parts of the moments) can be derived from the expression of the fields $\delta^{(p)}$.
Formally, Wick's theorem imposes~\footnote{this comes from a simple counting of lines: one needs at least $p-1$ lines to connect $p$ points which
in turns forces the sum of orders at which each factor is computed to be $2(p-1)$, see \cite{2002PhR...367....1B,1994A&A...291..697B,1992ApJ...392....1B,1984ApJ...279..499F}.} that the leading contributions 
to the $p$-order cumulant obtained from the following terms
\begin{equation}
\langle\rho^{p}\rangle_{c}=\sum_{\sum_{i=1}^{p} {n_{i}=2(p-1)}}
\langle\Pi_{i=1}^{p}\delta^{(n_{i})}\rangle_{c}\,.
\end{equation}
One of the well known consequences of that property is that $\langle\rho^{p}\rangle_{c}$ scales like $\langle\rho^{2}\rangle_{c}^{2(p-1)}$. It is then natural to  define precisely the reduced cumulants, $S_{p}$, as
\begin{equation}
S_{p}(\eta)=\frac{\langle\rho_{i}^{p}\rangle_{c}}{\langle\rho_{i}^{2}\rangle_{c}^{2(p-1)}}\,.
\end{equation}
It has been shown in \cite{1994A&A...291..697B,1992ApJ...392....1B} that these quantities are entirely determined by the dynamics of  the spherical collapse.  More precisely
the function $\zeta$ that relates the initial density contrast  $\tau_{<r}$ within a given shell of radius $r$ to the time-dependent ($\eta$)
non-linear density contrast,  $\rho_{<R}$ within the shell of radius $R=r\rho_{<R}^{-1/3}$,
\begin{equation}
\rho_{<R}=\zeta(\eta,\tau(<r))\,,
\label{zetadef}
\end{equation}
encodes all the necessary ingredients to compute the tree order cumulants. 
Note that the mere existence of such a function
takes full advantage of  Gauss' theorem, as the time evolution of the shell radius depends only on the density
contrast at this radius (before shell crossings).
More precisely,  if one perturbatively expands $\zeta(\eta,\tau)$ with respect to $\tau$,
\begin{equation}
\zeta(\eta,\tau)=\sum_{p}\nu_{p}(\eta)\frac{\tau^{p}}{p!}\,,
\end{equation}
(with $\nu_{0}=1$, $\nu_{1}=1$), then
each $S_{p}(\eta)$ parameters can be expressed in terms of $\nu_{p}(\eta)$.
For instance
\begin{eqnarray}
S_{3}(\eta)&=&3\nu_{2}(\eta)+\frac{\dd\log\langle\tau^{2}(r)\rangle}{\dd\log r},
\label{S3express}\\
S_4(\eta)&=&4\nu_3(\eta)+12\nu_2^2(\eta)+ \nonumber\\ 
&&
+\left(14\nu_2(\eta)-2\right){\dd\log[\langle\tau^{2}(r)\rangle]\over \dd\log r}+ \nonumber\\ 
&&\hskip -0.5cm +{7\over3}\left({\dd\log[\langle\tau^{2}(r)\rangle]\over \dd\log
r}\right)^2 +{2\over3}{\dd^2\log[\langle\tau^{2}(r)\rangle]\over \dd\log^2 r}.
\label{S4express}
\end{eqnarray}
The explicit form of $\zeta(\eta,\tau)$ or equivalently the values of $\nu_{p}(\eta)$ can a priori be predicted for any given cosmology. They  depend
on time --- although very weakly --- and take simple analytic forms for an Einstein de Sitter background. For instance, for such a background, we then have
$\nu_{2}=34/21$. A more general expression of $\zeta(\eta,\tau)$ can be found in \cite{1992ApJ...392....1B,1980lssu.book.....P,BernardeauBook}. 
In practice one can use a simple expression for $\zeta(\tau)$
\begin{equation}
\zeta(\tau)=\frac{1}{(1-\tau/\nu)^{\nu}}.
\label{zetaform}
\end{equation}
Here we  choose  $\nu=21/13\approx 1.6$  so that the high $z$ skewness 
of the density contrast  is exactly reproduced \footnote{We could choose $\nu=3/2$ in order to reproduce the low density asymptotic behavior of the exact solution.}. We checked that  this  choice of $\nu$  reproduce the exact spherical collapse dynamics for
Einstein-de Sitter background at a precision level of 0.5 \%  from $\zeta=0.3$ to $\zeta=2.5$ which is typically the range of values we need
to cover.

The understanding of the connexion between  the leading order statistical properties and the  spherical collapse dynamics has been dramatically
improved in \cite{1996A&A...312...11B,2000A&A...364....1B,2002A&A...382..412V} where it was realized that it  could be extended  to the cumulants of any number of concentric cells. We now turn to the presentation of
these results.

\subsection{General formalism}

We are here interested in the leading order expression of $\varphi=\varphi(\{\lambda_{k}\})$ for a finite number of concentric cells.
In this section we set the dimension of space to be $D$, having in mind that the formulae we derive should be valid for $D=2$ or $D=3$. For completeness, 
we sketch here the demonstration of the results and refer to  \cite{2002A&A...382..412V} for further details. To derive such an expression let us
introduce the joint density probability distribution functions,
$\mP(\{\rho_{k}\})\dd\rho_{1}\dots\dd\rho_{n}$, so that 
\begin{equation}
\exp\left[ \varphi(\{\lambda_{k}\})\right]=\int\dd\rho_{1}\dots\dd\rho_{n}
\mP(\{\rho_{k}\}) \exp(\sum_{i}\lambda_{i}\rho_{i}). \nonumber
\label{phiexp1}
\end{equation}
This expression can be written in terms of the  statistical properties of the {\sl initial} field. Let us define $\tau(\vx)$ as  the initial density contrast. Formally
the quantities $\rho_{i}$ are all functionals of the field, $\tau(\vx)$~\footnote{We implicitly assume throughout this paper that the initial conditions are adiabatic so that the perturbations developed out of a single scalar field degree of freedom.}, so that the ensemble average of the previous equation can be written as
\begin{equation}
\hskip -0.2 cm
\exp\left[ \varphi\right]\!\!=\!\! \int\mD\tau(\vx)\,\mP(\{\tau(\vx)\}) \exp(\sum_{i}\lambda_{i}\rho_{i}(\{\tau(\vx)\})),
\label{phiexp2}
\end{equation}
where we introduced the field distribution function, $\mP(\{\tau(\vx)\})$, and the corresponding measure $\mD(\{\tau(\vx)\})$. 
These are assumed to be known a priori. They depend on the initial conditions
and in the following we will assume the initial field is Gaussian distributed~\footnote{This is actually an superfluous hypothesis but we
adopt it for convenience. Transient effects
form mildly non-Gaussian initial conditions can for instance be taken into account in this formalism as shown in~ \cite{2002A&A...382..431V}.
We will briefly consider the case of non-Gaussian initial conditions in Subsect. \ref{sec:PNG}}. 

We now turn to the calculation of the generating function at leading order when the overall variance, $\sigma^2$, at scale $R_{i}$, is small. 
The idea is to identify the initial field
configurations that give the largest contribution to this integral. For convenience, 
let us assume that the field $\tau(\vx)$ can be described with a discrete number 
of variables $\tau_{i}$. For Gaussian initial conditions, the expression of the joint probability distribution function of $\tau_{i}$ reads
\begin{equation}
 \mP(\{\tau_{k}\})\dd\tau_{1}\dots\dd\tau_{p}=
\frac{\exp\left[-\Psi(\{\tau_{k}\})\right]}{\sqrt{(2\pi)^{p}/\det\Xi}}\dd\tau_{1}\dots\dd\tau_{p}\,,
\label{mPexp1}
\end{equation}
with
\begin{equation}
\Psi(\{\tau_{k}\})=\frac{1}{2}\sum_{ij}\Xi_{ij}\tau_{i}\tau_{j}\,,
\end{equation}
where $\Xi_{ij}$ is the {\sl inverse} of the covariance matrix, $\Sigma_{ij}$, defined as 
\begin{equation}
\Sigma_{ij}=\langle\tau_{i}\tau_{j}\rangle\,.
\label{sigmaij}
\end{equation}
The key idea to  transform Eq. (\ref{phiexp2}) using Eq.  (\ref{mPexp1})  relies on using the steepest descent
method. Details of the validity regime of this approach and its construction can be found in \cite{2002A&A...382..412V}.
The integral we are interested in is then dominated by a specific field configuration 
 for which the following stationary conditions are verified:
\begin{equation}
\sum_{i}\lambda_{i}\frac{\delta\rho_{i}(\{\tau_{k}\})}{\delta \tau_{j}}=\frac{\delta}{\delta \tau_{j}}\Psi(\{\tau_{k}\})\,,
\label{statCond1}
\end{equation}
for any value of $j$. Up to this point this is a very general construction.
Let us now propose a solution to these stationary equations that is consistent with
 the class of spherically symmetric problems we are interested in. The main point is the following: the configurations that are solutions of this equation, that is the values of $\{\tau_{k}\}$, depend specifically 
on the choice of the
functionals $\rho_{i}(\{\tau_{k}\})$. When these functionals correspond to spherically symmetric quantities, the corresponding configurations  are likely to be also spherically symmetric. But then  Gauss theorem is making things extremely simple: {\sl before shell crossing,} each of the the final density $\rho_{i}$
can indeed be expressed in terms of a {\sl single} initial quantity, namely the linear density contrast of the cell centered on $\vx_{0}$ 
that contained the  same amount of matter in the initial density field. We denote $\tau_{i}$ the corresponding density contrast,
which means that, following  definition (\ref{zetadef}), we have 
\begin{equation}
\rho_{i}=\zeta(\eta,\tau_{i}),\label{zetarelation}
\end{equation}
and  $\tau_{i}$ is the amplitude of the initial density within a specific radius~\footnote{This relation is a priori time dependent but we will 
omit it in the following.}, $r_{i}$, which obeys $r_i=R_{i}\rho_{i}^{1/D}$
thanks to mass conservation. 
The specificity of this mapping implies in particular that
\begin{equation}
\frac{\delta\rho_{i}(\{\tau_{k}\})}{\delta \tau_{j}}=\delta_{ij}\zeta'(\tau_{i})\,,
\end{equation}
so that the stationary conditions (\ref{statCond1}) now read
\begin{equation}
\lambda_{j}\zeta'(\tau_{j})=\frac{\delta}{\delta \tau_{j}}\Psi(\{\tau_{k}\})\,.
\label{statCond2}
\end{equation}
Note that the no-shell crossing conditions imply that if $R_{i}<R_{j}$, then $r_{i}<r_{j}$, which in turn 
implies that 
\begin{equation}
\rho_{i}<\rho_{j}(R_{j}/R_{i})^{D}.
\end{equation}
It follows that the parameter space $\{\rho_{k}\}$ is not fully accessible. In the specific example we  explore in the following,
this restriction is not significant, but it could be in some other cases.

We are now close to the requested expression for $\varphi(\{\lambda_{k}\})$ as we  have
\begin{equation}
\exp\left[ \varphi(\{\lambda_{k}\})\right]\!=\!\int\dd\tau_{1}\dots\dd\tau_{n}\!\mP(\{\tau_{k}\}) \exp(\sum_{i}\lambda_{i}\rho_{i}(\{\tau_{k}\}))\,.\nonumber
\end{equation}
To get the leading order expression of this form for  $\varphi(\{\lambda_{k}\})$, using the steepest descent method, one is simply requested to identify the quantities 
that are exponentiated. As a result we have
\begin{equation}
\varphi(\{\lambda_{k}\})=\sum_{i}\lambda_{i}\rho_{i}-\Psi(\{ \rho_{k}\})\,,\label{phifromLeg}
\end{equation}
where $\rho_{i}$ are determined by the stationary conditions (\ref{statCond2}). The latter
can be written equivalently as
\begin{eqnarray}
\lambda_{i}=\frac{\partial}{\partial \rho_{i}} \Psi(\{ \rho_{k}\})
\,,\label{statCond3}
\end{eqnarray}
when all quantities are expressed in terms of $\rho_{i}$.
Eq.~(\ref{statCond3}) is the general expression that we will exploit in the following. Formally, note that
 (\ref{phifromLeg})-(\ref{statCond2}) imply that $\varphi(\{\lambda_{k}\})$ is the Legendre transform
of $\Psi$ when the latter is seen as a function of $\rho_{i}$, that is
\begin{equation}
\Psi(\{ \rho_{k}\})=\frac{1}{2}\sum_{ij}\Xi_{ij}(\{ \rho_{k}\})\,\tau(\rho_{i})\tau( \rho_{j})\,,
\label{PsiDef}
\end{equation}
where
the functional form $\tau(\rho)$ is obtained from the inversion of (\ref{zetarelation}) at fixed time,
 and $\Xi_{ij}$ is the \textsl{inverse} matrix of the cross-correlation of the density in cells of radius
$R_{i}\rho_{i}^{1/D}$ (cf.  Eq.~(\ref{sigmaij}))
\begin{eqnarray}
\Sigma_{ij}&=&\left\langle\tau\left(<R_{i}\rho_{i}^{1/D}\right)\tau\left(<R_{j}\rho_{j}^{1/D}\right)\right\rangle\,,\\
\sum_{j}\Sigma_{ij}\,\Xi_{jk}&=&\delta_{ik}. \label{defsig}
\end{eqnarray}
These coefficients therefore depend on  the whole set of both radii $R_{i}$ and densities $\rho_{i}$. 
From the properties of Legendre transform, it follows in particular that  
\begin{eqnarray}
\rho_{i}=\frac{\partial}{\partial \lambda_{i}} \varphi(\{ \lambda_{k}\}).\label{statCond3bis}
\end{eqnarray}

Although known for more than a decade,  Eqs~(\ref{phifromLeg})-(\ref{statCond3bis})  and their consequences have not been exploited to their full power in the literature. This is partially
what we intent to do in this paper and in subsequent ones. For now, in order to get better acquainted with this formalism, let us first explore some of its properties.

\subsection{Scaling relations}
\label{Sub:scaling}

It is interesting to note that the cumulant generating function has a  simple dependence on
the overall amplitude of the correlators $\sigma^{2}_{0}$. Let us denote in  this subsection $\varphi_{\sigma_{0}}(\{\lambda_{k}\})$
the value of the cumulant generating function for a fixed value of $\sigma_{0}$. It is then straightforward to express 
$\varphi_{\sigma_{0}}(\{\lambda_{k}\})$ in terms of $\varphi_{1}(\{\lambda_{k}\})$, the expression of the generating function 
when $\sigma_{0}$ is set to unity. Indeed $\Psi(\{\rho_{k}\})$ is inversely proportional to $\sigma_{0}^{2}$
for fixed values of $\rho_{k}$. As a result $\lambda_{k}$ scale like $1/\sigma_{0}^{2}$ for fixed values of $\{\rho_{k}\}$. 
Note that we have the following identity
\begin{equation}\label{scaling}
\varphi_{\sigma_{0}}(\{\lambda_{k}\})=\frac{1}{\sigma_{0}^{2}}\varphi_{1}(\{\lambda_{k}/\sigma_{0}^{2}\}),
\end{equation}
while the variable $\rho_{k}$ are independent of $\sigma_{0}$.

In the upcoming applications  we will make use of this property as we will keep
the overall normalization as a free parameter -- that will eventually be adjusted on numerical results, but will
use the structural form of $\varphi_{1}(\{\lambda_{k}\})$ as predicted from the general theory. In particular this structural form
depends on the specific shape of the power spectrum through the cross-correlation matrix $\Sigma_{ij}$. 

\subsection{General initial conditions}
\label{sec:PNG}
The relation  Eqs~(\ref{phifromLeg})-(\ref{statCond3bis})   have been derived for Gaussian initial conditions. This eases the presentation but it is not 
a key assumption.  For instance in Eq. (\ref{mPexp1}), $\Psi(\{\tau_{k}\})$ does not need to be quadratic in $\tau_{k}$ as for Gaussian initial 
conditions. If the initial conditions were to
be non-Gaussian these features would have to be incorporated in the expression of $\Psi(\{\tau_{k}\})$.  
It would not however change the functional relation between $\varphi(\{\lambda_{k}\})$ and  $\Psi(\{\tau_{k}\})$, provided
$\Psi(\{\tau_{k}\})$ is properly defined when the variance is taken in its zero limit.

One would then observe that the Legendre transform between these two functions can be inverted~\footnote{This inversion is possible
as long as $\Psi(\{\tau_{k}\})$ is a convex function of $\rho_{k}$ which in practice means before shell crossing.}. 
Applying the fundamental relation  at precisely the initial time,  in a regime where $\rho_{i}\approx 1+\tau_{i} $, 
will give the expression of the function $\Psi(\{\tau_{k}\})$ in terms of the initial cumulant generating function. 
More precisely, if we define $\varphi_{\init}(\{\lambda_{k}\})$ as the cumulant generating function of the density contrasts 
at initial time, then we have
\begin{equation}
\Psi_{\init}(\{\tau_{k}\})=\sum_{i}\tau_{i}\lambda_{i}-\varphi_{\init}(\{\lambda_{k}\})\,,
\end{equation}
with
$
\tau_{i}=\partial \varphi_{\init} /{\partial \lambda_{i}}.
$
It is easy to check that for Gaussian initial conditions, 
\begin{equation}
\varphi_{\init}(\{\lambda_{i}\})=\frac{1}{2}\sum_{ij}\Sigma_{ij}\tau_{i}\tau_{j}\,,
\end{equation}
which leads to the expression (\ref{PsiDef}) given previously for $\Psi$. In this paper  we will however
use this construction for initial Gaussian conditions only.

\subsection{The one-cell generating function}

Turning back to application of Eqs~(\ref{phifromLeg})-(\ref{statCond3bis}),
one obvious simple application corresponds to the one cell characteristic function. In this case
\begin{equation}
\Psi(\rho)\equiv\frac{1}{2\sigma^{2}(R\rho^{1/D})}\tau(\rho)^{2}\,,
\end{equation}
where 
\begin{equation}
\sigma^{2}(r)=\langle \tau(<r)\tau(<r)\rangle.
\label{sigma2R}
\end{equation}
 The Legendre transform is then straightforward and  $\varphi(\lambda)$  takes the form
\begin{equation}
\varphi(\lambda)=\lambda\rho-\frac{1}{2\sigma^{2}(R\rho^{1/D})}\tau(\rho)^{2}\,,
\end{equation}
with $\rho$ computed implicitly as a function of $\lambda$ via  Eq.~(\ref{statCond3}). 
One way of rewriting this  equation is to define
$\eff\tau=\tau \sigma(R)/\sigma(R\rho^{1/3})$ and the function $\eff\zeta(\eff\tau)$
through the implicit form,
\begin{equation}
\eff\zeta(\eff\tau)=\zeta(\tau)=\zeta\left(\eff\tau \frac{\sigma(R \eff\zeta^{1/D})}{\sigma(R)}\right).
\end{equation}
Then the expression of $\varphi(\lambda)$ is given by
\begin{equation}
\varphi(\lambda)=\lambda\rho-\frac{1}{2\sigma^{2}(R)}\eff\tau^{2}\,,
\label{hzetaform1}
\end{equation}
with the stationary condition
\begin{equation}
\eff\tau=\lambda\sigma^{2}\eff\zeta'(\hat\tau).
\label{hzetaform2}
\end{equation}
In \cite{1994A&A...291..697B}, the expression of the cumulant generating function was presented with this form. 
This is also the functional form one gets when one neglects the filtering effects (as was initially done in \cite{1992ApJ...392....1B})
or for the so-called non-linear hierarchical model used in \cite{1992A&A...255....1B}.
Note that it is not possible however to use such a remapping  for more than one cell. 
Note finally  that this is a precious formulation for practical implementations,  as one may rely on fitted forms for $\eff\zeta$ to construct
the generating function $\varphi(\lambda)$ while preserving its analytical properties. 
It is indeed always possible, once one has been able to numerically compute $\varphi(\lambda)$ for specific values of
$\lambda$, to define $\eff\zeta$ by Legendre transform and construct fitted form with low order polynomials while
this is not possible for $\varphi(\lambda)$ which exhibits non-trivial analytical properties as we will see later on. This 
approach was used in \cite{2000A&A...364....1B}. It is also this procedure we use in Sect. \ref{Sect:profile} for constructing the profile PDF.

\subsection{Recovering the PDF via inverse Laplace transform}

In the following we will exploit the expression for the cumulant generating function to get the one-point and joint density PDFs.
To avoid confusion with the variables $\rho_{i} $ that appear in the expression of $\Psi$,
we will use the superscript $\hat{\,\,}$
to denote measurable densities, the PDF of which we wish to compute.

In general, the joint density PDF, $\mP=\mP(\hrho_{1},\dots,\hrho_{n})$, that gives the probability that the densities within 
a set of $n$ concentric cells of radii $R_{1},\dots R_{n}$ are $\hrho_{1}\dots\hrho_{n}$ within $\dd\hat\rho_{1}\dots\dd\hat\rho_{n}$
is given by
\begin{equation}
\mP=
\int_{-\ii\infty}^{+\ii\infty}\frac{\dd\lambda_{1}}{2\pi \ii}
\dots \frac{\dd\lambda_{n}}{2\pi \ii} \exp(-\sum_{i}\lambda_{i}\hrho_{i}+\varphi(\{\lambda_{k})\}). 
\label{InvLapTransnD}
\end{equation}
where the integration in $\lambda_{i}$ should be performed in the complex plane
so as to maximize convergence. 
This equation defines the inverse Laplace transform of the cumulant generating function
\footnote{It is beyond the scope of this paper to discuss the validity of this inversion and how it can 
generally  be implemented. Interested readers can find detailed discussions of this construction in \cite{1989A&A...220....1B}
in the context of cosmological density PDF and counts in cells statistics, where
it has been carried out in particular for the one-point PDF, the properties of which are crucially related
to the analytical properties of the cumulant generating function. }.
In the one-cell case, Eq.~(\ref{InvLapTransnD}) simply reads
\begin{equation}
\mP(\hrho_{1})=\int_{-\ii\infty}^{+\ii\infty}\frac{\dd\lambda_{1}}{2\pi \ii}\exp(-\lambda_{1}\hrho_{1}+\varphi(\lambda_{1}))\,,
\label{InvLapTrans1D}
\end{equation}
i.e. the PDF is  the inverse Laplace transform of the one-variable moment generating function. This inversion is known to be tricky,
and to our knowledge there are no known general full proof methods. One practical difficulty is that it generically relies
on the analytic continuation of the predicted cumulant generating function in the complex plane. It is therefore crucial to have  
a good knowledge of the analytic properties of $\varphi(\lambda)$, which is typically difficult since $\varphi(\lambda)$ is defined itself 
as the Legendre transform of $\Psi(\rho)$. Only a limited set of $\Psi(\rho)$ yield analytical $\varphi(\lambda)$, which in turn can be inverse-Laplace-transformed.

\section{The one-point PDF}
\label{Sect:OneCell}

{Up to this point,
the whole construction presented in the previous section would be a mere mathematical trick to compute explicit cumulants for top-hat window functions sparing the pain of lengthy integrations
on wave modes. 
In this paper, we furthermore aim to use the cumulant generating function computed
in the uniform limit $\Sigma_{ij}\to 0$ as an approximate form for the {\sl exact }generating function when the $\Sigma_{ij}$ are finite (but small). Note that this is a non-trivial extension for which we have no precise mathematical justifications. It assumes that the global properties of $\varphi(\{\lambda_{k}\})$
-- and in particular its analytical properties (which will be of crucial importance in the following), should be meaningful for {\sl finite values of $\lambda_{k}$}, and not only in the vicinity of $\{\lambda_{k}=0\}$. 

We now conjecture without further proof that
 they correctly represent the cumulant generating function {\sl for finite values of the variance}.}

\begin{figure}[ht]
\centerline{ \psfig{file=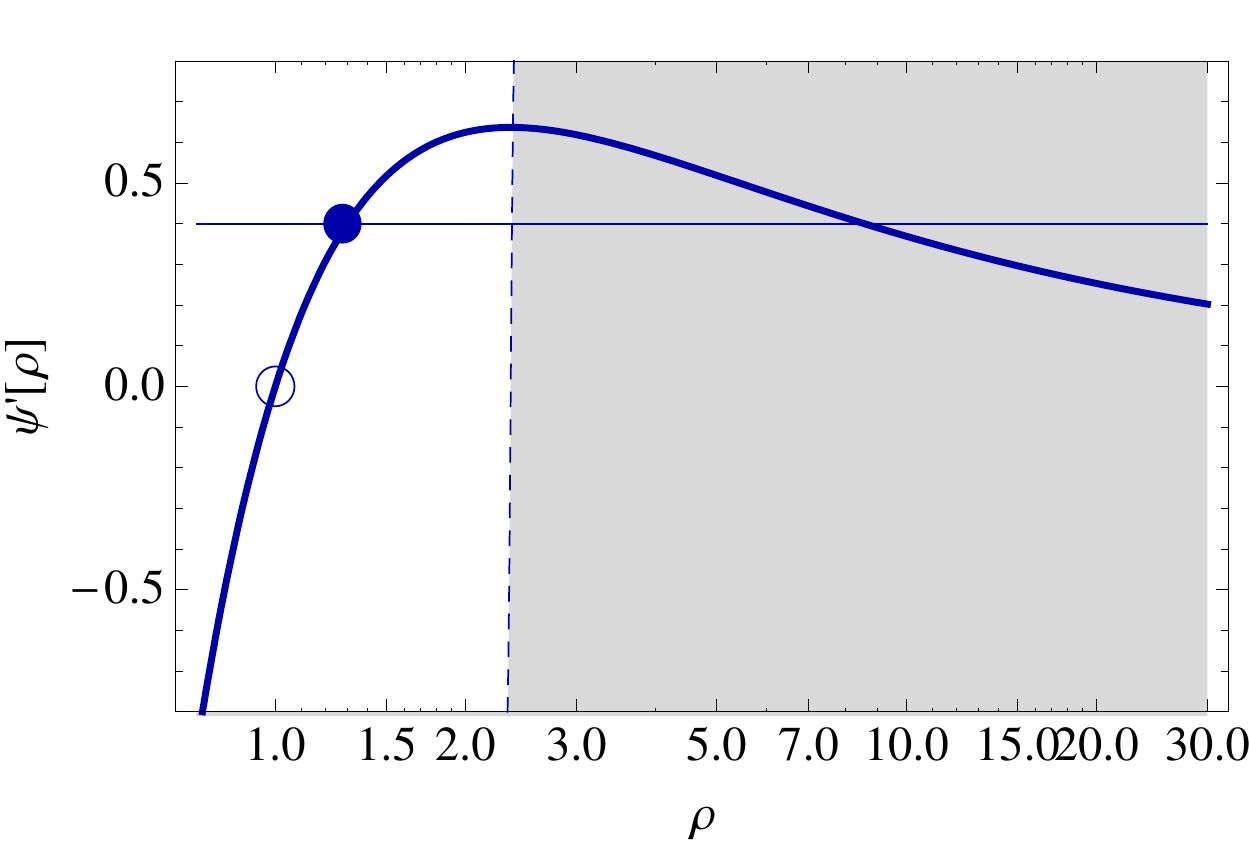, width=8 cm} }
   \caption{A graphical representation of the 1D stationary condition $\lambda=\Psi'[\rho]$. There is a maximum value for $\lambda$ that corresponds to a critical value $\rho_{c}$ for $\rho$  defined in Eq. (\ref{rhocdef}).}
      \label{1DCrit}
\end{figure}

\subsection{General formulae and asymptotic forms}
The implementation of the quadrature in Eq. (\ref{InvLapTrans1D}) has been attempted in various papers~\cite{1992A&A...255....1B,2000A&A...364....1B,1992ApJ...392....1B}, relying on
 different hypotheses for $\varphi(\lambda)$ \footnote{
In \cite{1992A&A...255....1B} this inversion was initially performed for models of nonlinear clustering (hierarchical models) taking advantage of the general results presented in \cite{1989A&A...220....1B}. The same techniques were later employed in the  
context of perturbation theory calculations. For all these constructions, the analytical properties of the cumulant generating function  follow from the structure of the stationary condition. }.
 Fig.~\ref{1DCrit} yields a graphical representation of the stationary equation for a power law model with index $n= -1.5$. 
The implicit equation, $\Psi'[\rho]=\lambda$, has always a solution in the vicinity of $\rho\approx 0$. Expanding this equation around this point naturally gives the low order cumulants at an arbitrary order.

Fig.~\ref{1DCrit} shows graphically that there is a maximum value for $\lambda$, $\lambda_{c}$, that can 
be reached, so that the Legendre Transform of $\Psi$ is not defined for $\lambda > \lambda_c$. It corresponds to a value $\rho=\rho_{c}$. At this location we have
\begin{equation}
0=\Psi''[\rho_{c}]\,,\qquad
\lambda_{c}=\Psi'[\rho_{c}] \,.\label{rhocdef}
\end{equation}
Note that at $\rho=\rho_{c}$, $\Psi$ is regular (in particular, the corresponding singular behavior in $\varphi(\lambda)$ is not related to any singularity of the spherical collapse dynamics).  The function $\varphi(\lambda)$ can  be expanded at this point.
I.e, Eq.~(\ref{statCond3})  can be inverted as a series near $(\rho_c,\lambda_c)$ (where Eq.~(\ref{rhocdef}) holds), and integrated for $\varphi(\lambda)$ using Eq.~(\ref{statCond3bis}). 
We give here a whole set of sub-leading terms that we will take advantage of in the following,
\begin{widetext}
\begin{eqnarray}
\varphi(\lambda)&=&
\varphi _{c}+\left(\lambda -\lambda _{c}\right) \rho _{c}+\frac{2}{3} \sqrt{\frac{2}{\pi _3}}
   \left(\lambda -\lambda _{c}\right)^{3/2}-\frac{\pi _4 \left(\lambda -\lambda _{c}\right)^2}{6 \pi
   _3^2} \nonumber\\
   &&\hspace{-1cm} 
   +\frac{\left(\frac{1}{\pi _3}\right){}^{7/2} \left(5 \pi _4^2-3 \pi _3 \pi _5\right) \left(\lambda
   -\lambda _{c}\right)^{5/2}}{45 \sqrt{2}}\   -\frac{\left(40 \pi _4^3-45 \pi _3 \pi _5 \pi _4+9 \pi _3^2 \pi
   _6\right) \left(\lambda -\lambda _{c}\right)^3}{810 \pi _3^5} \nonumber\\
&&\hspace{-1cm}   +\frac{\left(\frac{1}{\pi
   _3}\right){}^{13/2} \left(385 \pi _4^4-630 \pi _3 \pi _5 \pi _4^2+168 \pi _3^2 \pi _6 \pi _4+3 \pi _3^2
   \left(35 \pi _5^2-8 \pi _3 \pi _7\right)\right) \left(\lambda -\lambda _{c}\right)^{7/2}}{7560
   \sqrt{2}}+\dots
   \label{sphiexp}
\end{eqnarray}
\end{widetext}
where $\pi_{n}={\partial^{n}\Psi}/{\partial\rho^{n}}(\rho_{c})$. 
 It is to be noted that the leading singular term scales like
 $\left(\lambda -\lambda _{c}\right)^{3/2}$. The coefficients $\pi_{i}$ are all related to the function $\Psi$ and are therefore (cosmological)
 model dependent~\footnote{Note that in practice $\pi_{3}$ is negative and that only the non regular parts appearing 
in the expansion of $\exp(\varphi(\lambda))$ will contribute.}.

What are the consequences of this behavior for the  PDF of the density? Let us present analytical forms for the inverse Laplace transform of $\exp \varphi$. The idea is that the inverse transform can be obtained  via a saddle point approximation of Eq.~(\ref{InvLapTrans1D}) assuming the variance is small. Formally
it leads to the conditions that should be met at the saddle point $\lambda_{s}$, 
\begin{eqnarray}
\frac{\partial}{\partial \lambda}\left[\lambda\hrho-\varphi(\lambda)\right]&=&0\,,\\
\frac{\partial^{2}}{\partial \lambda^{2}}\left[\lambda\hrho-\varphi(\lambda)\right]&<&0.
\end{eqnarray}
The first condition leads to $\rho(\lambda_{s})=\hrho$, while  the second to $\lambda_{s}<\lambda_{c}$. This condition simply means that this approximation
can be used if $\hrho<\rho_{c}$. 
The resulting simple expression for the density PDF is 
\begin{equation}
\mP(\hrho)=\frac{1}{\sqrt{2 \pi}}
\sqrt{\frac{\partial^{2}\Psi(\hrho)}{\partial\hat\rho^{2}}}\exp\left[-\Psi(\hrho)\right].
\label{aPDFlowrho}
\end{equation}
It is valid as long as the  expression that appears in the square root is positive, {\it i.e.}
$\hrho<\rho_{c}$. 
When this  condition is not satisfied, the singular behavior of $\varphi$ near $\lambda_c$ dominates the integral in the complex plane.
This leads to 
the following expression for $\mP(\hrho)$ as described in appendix \ref{tails},
\begin{widetext}
\begin{equation}
\mP(\hrho)\approx
\exp\left(\varphi _{c}- \lambda _{c}\hrho \right)
\left(\frac{3\, \Im{(a_{\frac{3}{2}} )} }{4 \sqrt{\pi } \left(\hrho -\rho _{c}\right)^{5/2}}+\frac{15\, \Im(a_{\frac{5}{2}})}{8
   \sqrt{\pi } \left(\hrho -\rho _{c}\right)^{7/2}}+\frac{105 \left(\Im(a_{\frac{3}{2}})
   a_2+\Im(a_{\frac{7}{2}})\right)}{16 \sqrt{\pi } \left(\hrho -\rho _{c}\right)^{9/2}}+\dots \right) \,,
   \label{aPDFlargerho1}
\end{equation}
\end{widetext}
where $a_{j}$ are the coefficients in front of $\left(\lambda -\lambda _{c}\right)^{j}$ in Eq.~(\ref{sphiexp}), (e.g. 
$a_{3/2}={2}/{3} \sqrt{{2}/{\pi _3}}$) and $\Im(\,\,)$ is the imaginary part.
Eq.~(\ref{aPDFlargerho1}) has an exponential cut-off at large $\hrho$ scaling like $\exp(\lambda_{c}\hrho)$.  This property is actually robust and is preserved 
when one performs the inverse Laplace transform for finite values of the variance, or even large value of the variance (see \cite{1989A&A...220....1B,1992ApJ...390L..61B}).
It also gives a direct transcription of why $\varphi(\lambda)$ becomes singular: for values of $\lambda$ that are larger than $\lambda_{c}$, the integral
$\int\!\dd\hrho\, \mP(\hrho) \exp(\lambda\hrho)$ is not converging.

Note that in practice, it is best to rely on an alternative asymptotic form to Eq. (\ref{aPDFlargerho1})  that  is better behaved and remains finite for
$\hrho\to \rho_{c}$. It is built in such a way that it has the same asymptotic behavior as Eq.~(\ref{aPDFlargerho1}) at a given order in the large $\rho$ limit. The following form
\begin{equation}
\mP(\hrho)=
\frac{3 a_{\frac{3}{2}} \exp\left(\varphi _{c}- \lambda _{c}\hrho \right)
}{4 \sqrt{\pi }\left(\hrho +r_{1}+r_{2}/\hrho+\dots \right)^{5/2}},
   \label{aPDFlargerho2}
\end{equation}
where the $r_{i}$ parameters are adjusted to fit the results of the previous expansion, proved very robust. At NLO and NNLO we have
\begin{eqnarray}
r_{1}&=&-\frac{\Im(a_{\frac{5}{2}})}{\Im(a_{\frac{3}{2}})}-\rho _{c},\\
r_{2}&=&-\frac{7 \left(2 a_2 a_{\frac{3}{2}}^2+2 a_{\frac{7}{2}} a_{\frac{3}{2}}-a_{\frac{5}{2}}^2\right)}{4
   a_{\frac{3}{2}}^2}.
\end{eqnarray}
\begin{figure*}[ht]
\centerline{ \psfig{file=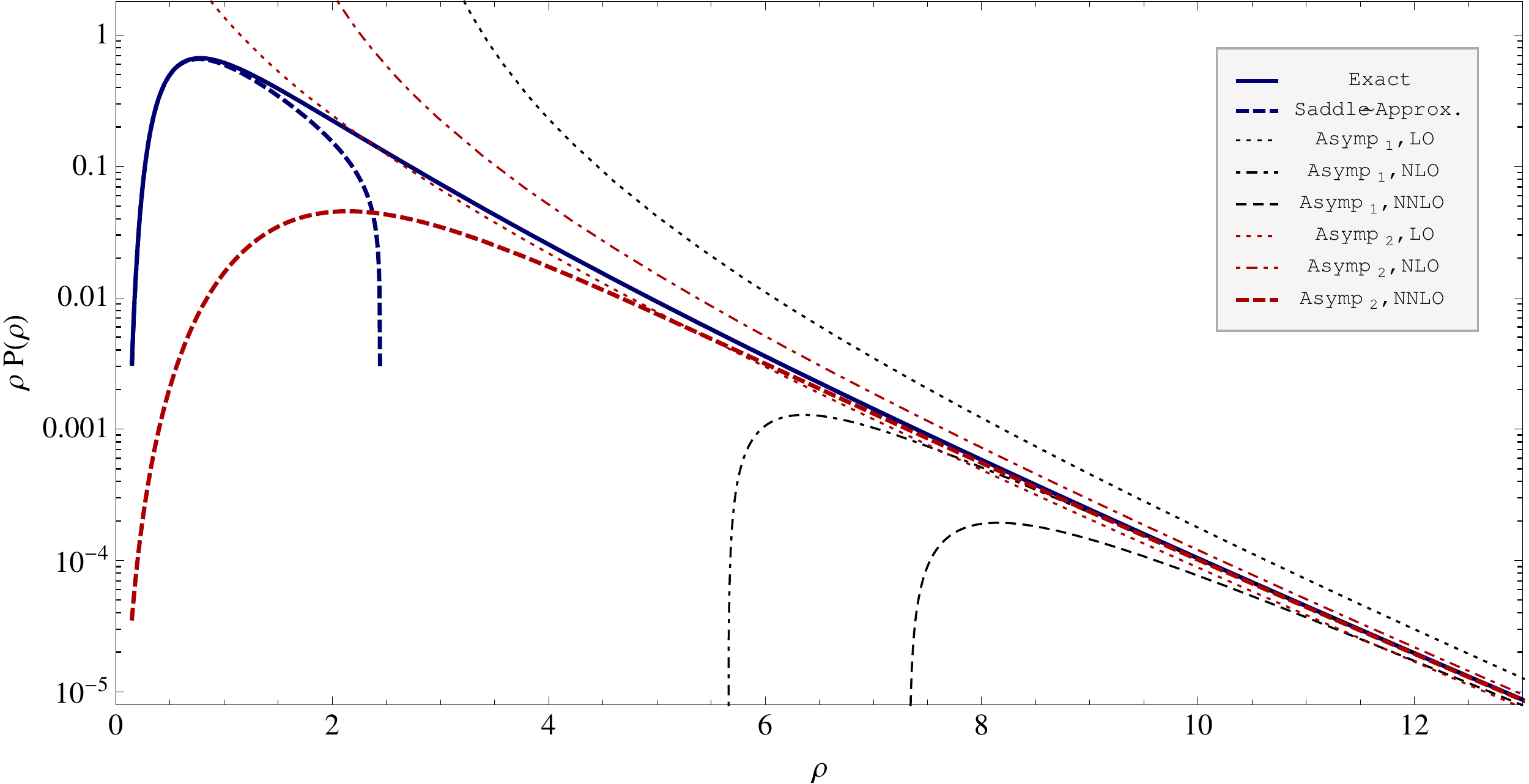, width=16 cm} }
   \caption{The PDF of the one-point density. The blue solid line is the numerical integration; the red dashed line the low $\rho$ asymptotic form of Eq. (\ref{aPDFlowrho}); the other lines correspond to the large $\rho$ asymptotic forms proposed in the text: The dark lines correspond to the form  (\ref{aPDFlargerho1})  and the red lines to the form  (\ref{aPDFlargerho2})  and  the forms are computed at leading order, at next-to-leading, and next-to-next-to-leading order for respectively the dotted, dot-dashed and dashed curves. The plot are given for $\sigma^{2}=0 .45$ and   a power law index of $n=-1.5$.
   \label{Prho1D}}
\end{figure*}
However, none of these asymptotic forms  are accurate for the full range of density values; in general one has to rely on numerical integrations
in the complex plane which can be done accurately and quickly, as described in appendix \ref{Integrationinthecomplexplane}.
The comparison between the analytical forms
and the numerical integrations are shown in Fig. \ref{Prho1D}. Such comparisons are in fact conversely useful to assess the precision of the numerical integrations.
Note that for the case explicitly shown, which corresponds to $\sigma^{2}= 0.45$ and 
a power law index of $n=-1.5$, the asymptotic forms (\ref{aPDFlowrho}) and (\ref{aPDFlargerho2}) at NNLO are valid within 2 \%
everywhere but for the range $1<\rho <10$, where one must rely on an explicit integration in the complex plane.

\subsection{Practical implementation, comparisons with N-body results}

We now move to an explicit comparison of these predictions to N-body results. The simulations are described in
appendix \ref{Simulations}. They are determined in particular by the linear power spectrum $P^{\lin}(k)$
set for the initial conditions.  The knowledge of the power spectrum determines the values of the cross-correlation
matrix, $\Sigma_{ij}(R_{i},R_{j})$, that are explicitly given by
\begin{equation}
\Sigma_{ij}=\int\frac{\dd^{3}\vk}{(2\pi)^{3}}P^{\lin}(k)W_{\rm 3D}(k R_{i})W_{\rm 3D}(k R_{j})\,,
\end{equation}
 where $W_{\rm 3D}(k)$ is the shape of the top-hat window function in Fourier space,
 \begin{equation}
W_{\rm 3D}(k)=3\sqrt{\frac{\pi}{2}}\frac{J_{3/2}(k)}{k^{3/2}}\,
\end{equation}
where $J_{3/2}(k)$ is the Bessel function of the first kind of index $3/2$. In 3D, it is actually possible to 
express $W_{\rm 3D}(k)$ in terms of elementary functions as
\begin{equation}
W_{\rm 3D}(k)=\frac{3}{k^{2}}\left(\sin(k)/k-\cos(k)\right).
\end{equation}
For the one-cell case we only need to know the amplitude and scale dependence of 
$\sigma^{2}_{R}$ defined as
\begin{equation}
\sigma^{2}(R)=\int\frac{\dd^{3}\vk}{(2\pi)^{3}}P^{\lin }(k)W^{2}_{\rm 3D}(k R).
\end{equation}

To a first approximation, $\sigma^{2}(R)$ can be parametrized with a simple power law $\sigma^{2}(R)\sim R^{-(n_{s}+3)}$.
It is this functional form which was used in the previous section. The detailed predictions of the PDF depend however
on the precise scale dependence of $\sigma^{2}(R)$. Such scale dependence can be computed numerically from the 
shape of the power spectrum but it makes then difficult to derive the function $\varphi(\lambda)$ from Legendre transform. So
in order to retain simple analytic expressions for the
whole cumulant generating function, we adopt a simple prescription for the scale dependence
of $\sigma^{2}(R)$ given by
\begin{equation}
\sigma^{2}(R)=\frac{2\sigma^2(R_{p})}{(R/R_p)^{n_1+3}+(R/R_p)^{n_2+3}}\,,
\end{equation}
where $R_{p}$ is a pivot scale. Such a parametrization
ensures that the single-point $\Psi(\rho)$ function takes a simple {\sl analytic} form as it involves the {\sl inverse} of $\sigma^{2}(R)$.
Note that our Ansatz can be extended to an arbitrary (finite) number of terms in the denominator.

The values of the three parameters, $\sigma^2(R_{p})$, $n_{1}$ and $n_{2}$ are then adjusted so that the model reproduces i) the \textsl{measured}
variance $\sigma^{2}(R)$,  
ii) the linear theory index
\begin{equation}
n(R)=-3-\frac{\dd\log(\sigma(R))}{\dd\log R}\,,
\end{equation}
and iii) its running parameter
\begin{equation}
\alpha(R)=\frac{\dd\log(n(R))}{\dd\log R}\,,
\end{equation}
at the chosen filtering scale. 
%
It is important to point out that we do not take the amplitude of $\sigma^{2}(R)$ as predicted
by  linear theory. We consider instead its overall amplitude as a free parameter and 
$\sigma^{2}(R)$ is directly measured from the N-body results. The reason is that using the predicted
value of $\sigma^{2}(R)$ would simply introduce too large errors  and this dependence can always be scaled out
using the relation of Sect. \ref{Sub:scaling}~\footnote{An alternative approach 
would be to use predicted amplitude form direct next-to-leading order perturbation theory
calculations. We leave this option for further studies.}.

\begin{figure}[ht]
\centerline{ \psfig{file=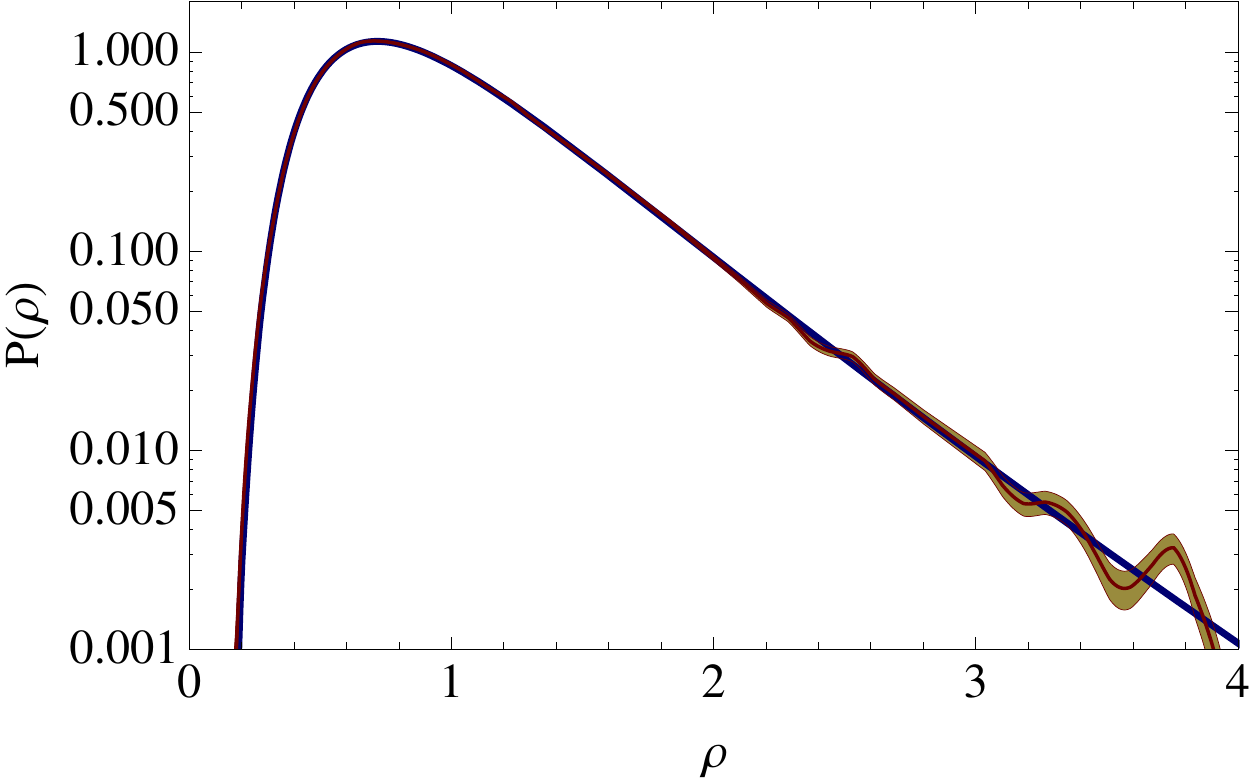, width=8 cm} }
\centerline{ \psfig{file=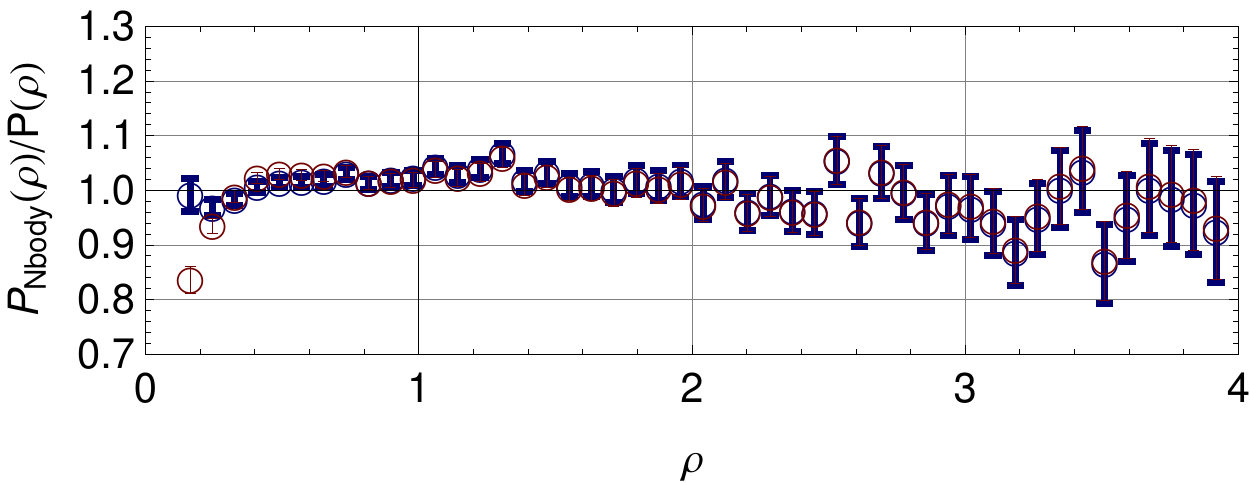, width=7.8 cm} }
   \caption{Comparison with simulations (top) with residuals (bottom). The solid line is the theoretical prediction
    computed  for a variance of $\sigma^{2}_{R}=0 .47$ as measured in the simulation,  
   a power law index of $n=-1.576$ and a running parameter $\alpha=0.439$ 
   corresponding to the input linear power spectrum. The measured PDF in the simulation is shown as 
   a band corresponding to its 1--$\sigma$ error bar (but different data points are correlated).
   The residuals show the ratio of measured PDF in bins with the predictions (computed in bins as well). The thin red
   symbols show the comparison when the running parameter is set  to zero in the prediction.}
\label{ExpPDF10}
\end{figure}

In Fig. \ref{ExpPDF10}, we explicitly show the comparison between our predictions following the prescription we just described 
to measured PDFs. The predictions show a remarkable agreement with the measured PDF! 
Recall that only one parameter, $\sigma_{R}$, is adjusted to the numerical data. In particular the predictions reproduce with a extremely  good accuracy the PDF tails in both the low density and high density regions. 
The plot of the residuals shows the predictions are at the percent level over a large range of
density values. And this result is obtained for a squared variance  close to $0.5$.

More extended comparisons  with numerical simulations are shown on Fig. \ref{ExpPDF-zp97-rest} 
which qualifies in more details  the validity regime of our predictions. Note that 
up to $\sigma=0.64$ ($\sigma^{2}=0.41$), we see no significant departure from the results of the simulation in the whole range of available 
densities, that is in particular up to 
about the $5\sigma$ rare event in the high density tail. This success is to be contrasted with the Edgeworth expansion approach
which breaks for $\vert\delta\vert\ge\sigma$ (see for instance \cite{1995ApJ...443..479B}).

We observe that departures from our calculations start to be significant,  of the order of $10\%$ , when $\sigma^{2}(R)$
is of the order of $0.7$ or more \footnote{The difference can be, to a large extent, interpreted by the nonlinear growth of the reduced
skewness. }.
%
%
\begin{figure}[t]
\centerline{ \psfig{file=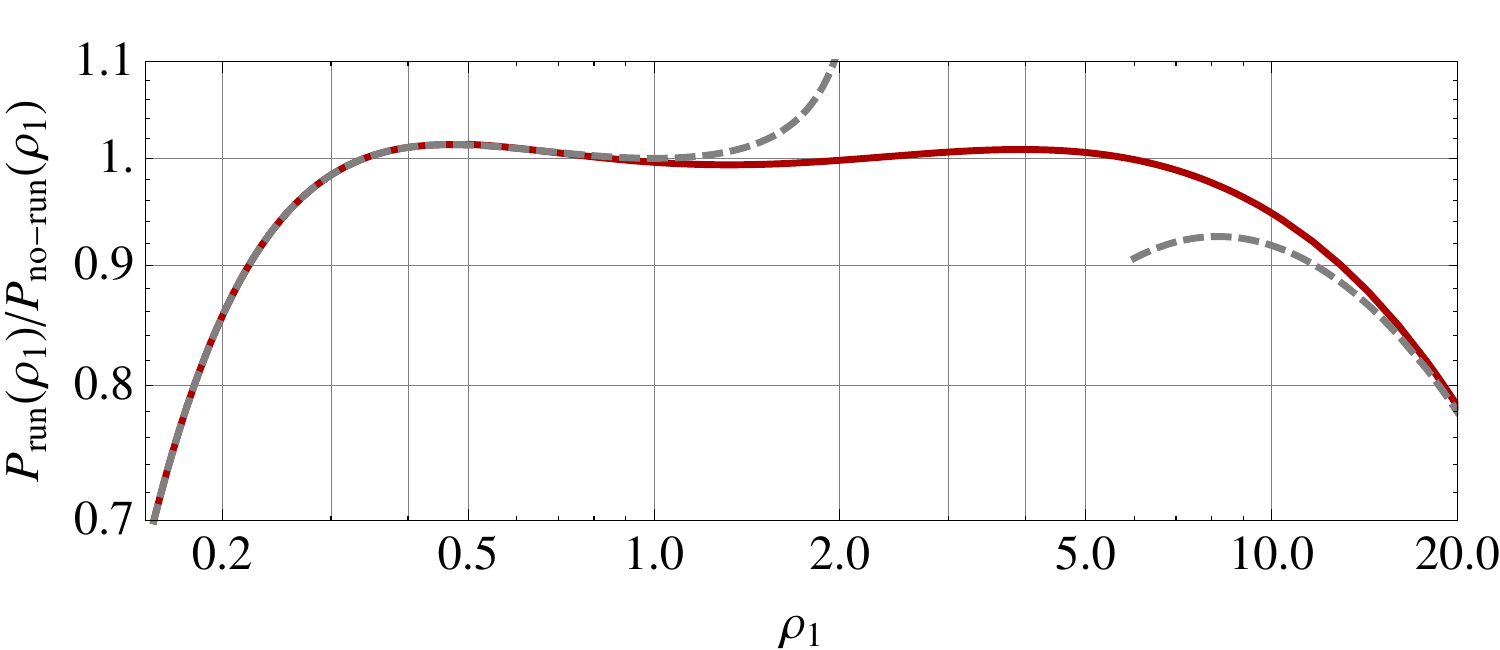, width=8 cm} }
   \caption{Ratio of the one-point density PDF when the running parameter is taken into account over the PDF when it is not.  
   The running model
   is the same as in the previous plot. The dashed lines are the ratio of the corresponding asymptotic forms in the low and
   high density regions.
   \label{DiffPrho1D}}
\end{figure}
These results also show that taking into account the scale dependence of the local index through the introduction
of the running parameter improves upon the predictions in the low density region. This aspect is examined in more detail in 
Fig. \ref{DiffPrho1D} which shows the ratio of the predicted PDFs with and without taking into account the running parameter.
We see that  the PDFs are mostly affected on their tails. This is related to the fact that the kurtosis is the lowest order 
cumulant to be changed  when one introduces a running parameter~\cite{1994ApJ...433....1B}, as can be verified from the relation (\ref{S4express}). The effect is actually detectable in the low density region only
and confirms the fact that the introduction of a running parameter
can have a noticeable impact when comparisons at the percent level are to be done.

\section{The statistical properties of the density slope and profile}
\label{Sect:profile}

We now move to the application of the general formalism to the two-cells case. Such situations have already been encountered in \cite{1996A&A...312...11B} to compute effective bias properties, and in \cite{2000A&A...364....1B} to compute the aperture mass statistics
out of two concentric angular cells of fixed radius ratio. But all these applications eventually reduce to an effective one-cell case.
We are interested here in genuine two-cell statistics.

Let us first make a remark that may seem trivial. Indeed, from the very definition
of cumulant generating functions, one should have
\begin{equation}
\varphi_{2-\cell}(\lambda_{1},\lambda_{2}=0)=\varphi_{1-\cell}(\lambda_{1})\,,
\end{equation}
where $\varphi_{2-\cell}(\lambda_{1},\lambda_{2})$ is the cumulant generating function for cells of radii $R_{1}$ and $R_{2}$
and $\varphi_{1-\cell}(\lambda_{1})$ is the cumulant generating function for one cell of radius $R_{1}$.
Checking that the relations (\ref{phifromLeg}--\ref{statCond3bis}) verify this property makes a sound mathematical exercise! More generally 
one can show that  our formulation is consistent with radii decimation, that is when one computes
the cumulant generating functions of a restricted number of variables out of a larger number one gets a consistent result.   
The demonstration of this property is given  in Appendix \ref{decimation}.

The purpose of this section is  now to define the statistical properties of the density profile, while relying on the fact that the function $\varphi(\lambda_{1},\lambda_{2})$ has a well defined, but non trivial limit,
when one sets $\Delta R=R_{2}-R_{1}\ll R_{1}$.

\subsection{The density slope}

From the densities in two concentric cells, it is indeed always possible to define the corresponding density \emph{slope} as
\begin{equation}
\hs(R_{1},R_{2})=\frac{R_{1}}{\Delta R}\left[\hrho_{2}-\hrho_{1}\right].
\label{slopedef}
\end{equation}
In the limit of a vanishing smoothing radius difference, $\hs$ will define the local density slope. 
In the following we will in particular see that this is a genuine limit in the sense that it leads to regular and non trivial expressions. 

Let us start with basic preliminary calculations;  to avoid too complicated notations,  let us define
\begin{eqnarray}
\sigma^{2}_{R_1}&\equiv&\sigma^{2}(R_{1},R_{1}),\\
\sigma^{2}_{R_1R_2}&\equiv&\sigma^{2}(R_{1},R_{2}),\\
\sigma^{2}_{R_2}&\equiv&\sigma^{2}(R_{2},R_{2}),
\end{eqnarray}
which are quantities involved in the expressions of cumulants.
The variance of $\hat s$ is then for instance given by
\begin{equation}
\langle \hs^{2}\rangle=\left(\frac{R_{1}}{{\Delta R}}\right)^{2}(\sigma^{2}_{R_{1}}-2 \sigma^{2}_{R_{1}R_{2}}+\sigma^{2}_{R_{2}})\,.
\end{equation}

From the general theory, Eqs.~(\ref{phifromLeg})-(\ref{statCond3bis}) implemented for two cells, one can compute the generating function of joint density contrasts in concentric cells \footnote{In practice this is achieved via a series expansion inversion of Eq.~(\ref{statCond3})
then plugged in Eq.~(\ref{phifromLeg}).} in the limit of small $\lambda_i$. 
Up to third order it is explicitly given by
\begin{widetext}
\begin{eqnarray}
&&\varphi(\lambda_{1},\lambda_{2})=\lambda_{1}+\lambda_{2}+
\frac{1}{2} \lambda _1^2\ \sigma^2_{R_1}\frac{}{}
+\frac{1}{2} \lambda _2^2\ \sigma^2_{R_2}+\lambda _1 \lambda _2\ \sigma^2_{R_1R_2}\nonumber\\
&&
 +\lambda _1^3 \left(\frac{1}{2} \nu _2 \sigma^4_{R_1}+\frac{1}{6} R_1 \sigma^2_{R_1} \frac{\dd}{\dd R_1}\sigma^2_{R_1}\right)+\lambda _2^3 \left(\frac{1}{2} \nu _2 \sigma^4_{R_2}+\frac{1}{6} R_2 \sigma^2_{R_2} \frac{\dd}{\dd R_2}\sigma^2_{R_2}\right)
\nonumber\\
&&+\lambda _1^2 \lambda _2 \left(\frac{1}{2} \nu _2 \sigma^2_{R_1R_2} \left(\sigma^2_{R_1,R_2}+2 \sigma^2_{R_1}\right)+\frac{1}{6} \left(2
   R_1 \sigma^2_{R_1} \frac{\partial}{\partial R_1}\sigma^2_{R_1R_2}+\sigma^2_{R_1R_2} \left(2 R_2 \frac{\partial}{\partial R_2}\sigma^2_{R_1R_2}+R_1 \frac{\dd}{\dd R_1}\sigma^2_{R_1}\right)\right)\right)
  \nonumber\\
  &&+ \lambda _1 \lambda _2^2 \left(\frac{1}{2} \nu _2 \sigma^2_{R_{1}R_2}
   \left(\sigma^2_{R_1R_2}+2 \sigma^2_{R_2}\right)+\frac{1}{6} \left(2 R_2 \sigma^2_{R_2} \frac{\partial}{\partial R_2}\sigma^2_{R_1R_2}+\sigma^2_{R_1R_2} \left\{2 R_1 \frac{\partial}{\partial R_1}\sigma^2_{R_1R_2}+R_2 \frac{\dd}{\dd R_2}\sigma^2_{R_2}\right\}\right)\right)\,,
 \label{eqvarphiupto3}
 \end{eqnarray}
\end{widetext}
where $\nu_{2}=34/21$ for a 3D dynamics in an Einstein de Sitter background.
In Eq.~(\ref{eqvarphiupto3}), the  cumulants and joint cumulants can be read out using  definition (\ref{phidef}) or via differentiation.
For instance,
\begin{eqnarray}
\langle \hrho_{1}^{3}\rangle_{c}&=&3\nu _2 \sigma^4_{R_1}+\sigma^2_{R_1} \frac{R_{1}\dd}{\ \dd R_1}\sigma^2_{R_1}\,,\\
\langle \hrho_{1}^{2}\hrho_{2}\rangle_{c}&=&
\nu _2 \sigma^2_{R_1R_2} \left(\sigma^2_{R_1 R_2}+2 \sigma^2_{R_1}\right)\nonumber\\
&+&\frac{2}{3}  \sigma^2_{R_1} \frac{R_{1}\partial}{\partial R_1}\sigma^2_{R_1R_2}\nonumber\\
&+&\frac{1}{3}\sigma^2_{R_1R_2} \left(2  \frac{R_2\partial}{\partial R_2}\sigma^2_{R_1R_2}+\frac{R_{1}\dd}{\ \dd R_1}\sigma^2_{R_1}\right),
\end{eqnarray}
and the cumulants $\langle\rho_{1}\rho_{2}^{2}\rangle_{c}$ and $\langle\rho_{2}^{3}\rangle_{c}$ can be obtained exchanging the role of $R_{1}$
and $R_{2}$.
It is then also possible to derive the explicit form for a number of auto- and cross-cumulants between the density $\hrho\equiv\hrho_{1}$ in the first cell
and the slope $\hs$ as defined in (\ref{slopedef}). 
For instance,
\begin{eqnarray}
\langle \hrho^{2}\hs\rangle_{c}&=&\frac{R_{1}}{\Delta R}\left[\langle \hrho_{1}^{2} \hrho_{2}\rangle_{c}-\langle \hrho_{1}^{3}\rangle_{c}\right]\,,\\
\langle \hrho\, \hs^{2}\rangle_{c}&=&\left[\frac{R_{1}}{\Delta R}\right]^{2}\left[\langle \hrho_{1}\hrho_{2}^{2}\rangle_{c}-2\langle \hrho_{1}^{2} \hrho_{2}\rangle_{c}+
\langle \hrho_{1}^{3}\rangle_{c}
\right]\,,\\
\langle \hs^{3}\rangle_{c}&=&\left[\frac{R_{1}}{\Delta R}\right]^{3}\nonumber\\
&\times& \hskip -0.2cm 
\left[
\langle \hrho_{2}^{3}\rangle_{c}
-3\langle \hrho_{1} \hrho_{2}^{2}\rangle_{c}
+3\langle \hrho_{1}^{2} \hrho_{2}\rangle_{c}
-\langle \hrho_{1}^{3}\rangle_{c}
\right].
\end{eqnarray}
Following the one cell case (see for instance  \cite{2002PhR...367....1B})
 it is possible to formally define the reduced cross-correlations that are independent on the overall amplitude of the power
spectrum. More precisely,  the reduced cross-correlations can be defined  as
\begin{eqnarray}
S_{p0}&=&\frac{\langle\hrho^{p}\rangle_{c}}{\langle\hrho^{2}\rangle_{c}^{p-1}}\,,\\
S_{pq}&=&\frac{\langle\hrho^{p}\hs^{q}\rangle_{c}}{\langle\hrho^{2}\rangle_{c}^{p-1}\langle \hrho \hs\rangle_{c}\langle \hs^{2}\rangle_{c}^{q-1}}\,,\\
S_{0q}&=&\frac{\langle \hs^{q}\rangle_{c}}{\langle \hs^{2}\rangle_{c}^{q-1}}\,.
\end{eqnarray}
From the previous expressions these quantities can be computed in the limit of an infinitely small variance.

\subsection{Cumulants and  slope in the limit $(\Delta R)/R\to 0 $}

Let us now consider the statistical properties of $\hs$ in the limit $(\Delta R)/R\to 0 $.
To start with, let us compute the variance of the slope $\hs$ in the limit $\Dr/R\to 0$.
Its variance is formally  given by
\begin{equation}
\langle \hs^{2} \rangle=\left.{\frac{R_{1}^{2}\ \partial^{2}}{\partial R_{1} \partial R_{2}}\sigma^{2}_{R_{1}R_{2}}}\right\vert_{R=R_{1}=R_{2}}.
\end{equation}
This expression can easily be expressed in terms of the
power spectrum,
\begin{equation}
\langle\hs^{2}\rangle=\int\frac{\dd^{3}\vk}{(2\pi)^{3}}P^{\lin}(k)\tW_{\rm 3D}^{2}(kR),
\end{equation}
where $\tW_{\rm 3D}(k)$ is the logarithmic derivative of $W_{\rm 3D}(k)$,
\begin{equation}
\tW_{\rm 3D}(k)=\frac{\dd}{\dd \log k}W_{\rm 3D}(k),
\end{equation}
which for the 3D case can be written,
\begin{equation}
\tW_{\rm 3D}(k)=\frac{1}{k^{3}}\left[(9 k \cos(k) + 3 (k^2-3) \sin(k)\right].
\end{equation}
Note that for a power law spectrum of index $n_{s}$ this variance is only defined when $n_{s}<-1$. For practical application to cosmological models that resemble the concordant model, the effective index $n_{s}$ decreases to $-3$ at small scales 
and the variance of $\hs$ is always finite. This property however suggests that the
amplitude of the slope fluctuations could be dominated by density fluctuations at scales significantly smaller
than the smoothing radius if the latter is large enough. 
This is not expected to be the case however for the filtering scales we explore in this investigation.
More precisely, provided the power spectrum index is in the range $[-3,-1]$, the amplitude
of the variance of $\hs$ can be expressed in terms of the variance of the density as,
\begin{equation}
\langle\hs^{2}\rangle=\sigma^{2}_{R}\ \frac{n_{s} (n_{s}+3) (n_{s}+5)}{4 (n_{s}+1)}.
\end{equation}

Let us now  see how the whole statistical properties of the variable $\hs$ can be derived from our formalism.
Let us  first explore the  consequence of the change of variable, $(\hrho_{1},\hrho_{2})\to (\hrho,\hs)$.
Instead of describing the joint PDF as a function of the associated variables $\lambda_{1}$ and $\lambda_{2}$
we can build it with the variable associated to $\hrho$ and $\hs$. Noting that $\lambda_{1}\hrho_{1}+\lambda_{2}\hrho_{2}$ can be written as
\begin{equation}
\lambda_{1}\hrho_{1}+\lambda_{2}\hrho_{2}=(\lambda_{1}+\lambda_{2})\hrho_{1}+\frac{\Delta R}{R_{1}}\lambda_{2}\hs\,,
\end{equation}
as a consequence, the joint cumulant generating function of $\hrho_{1}$ and $\hs$ is 
given by $\varphi(\lambda_{1},\lambda_{2})$ when written as a function of 
\begin{equation}
\lambda=\lambda_{1}+\lambda_{2}\,,\quad
\mu=\frac{\Delta R}{R_{1}}\lambda_{2},
\end{equation}
which are  the variables associated with the Laplace and inverse Laplace transform of $\mP(\hrho_{1},\hs)$. One can also 
 check that,  following this definition, $\varphi(\lambda,\mu)$ is the Legendre transform of $\Psi(\rho_{1},s=(\rho_{2}-\rho_{1})\,R_{1}/\Delta R)$.

Let us then explore the whole statistical properties of $\hs$ in the limit of a vanishing radius difference $(\Delta R)/R\to 0$.
First  note that the reduced skewness of $\hs$ is still finite \footnote{again when $n_{s}$ is less than -1 in case of a power law spectrum.} and has a non trivial value. It is given by
\begin{equation}
S_{03}^{\Delta R\to 0}=2+\left.\frac{\frac{\partial}{\partial R_{1}}\sigma^{2}_{R_{1}R_{2}}}{\frac{R_{1}\partial ^{2}}{\partial R_{1} \partial R_{2}}\sigma^{2}_{R_{1}R_{2}}}\right\vert_{R=R_{1}=R_{2}}\hskip -0.5cm \left(6\nu_{2}-(\tilde{n}+3)\right),
\end{equation}
where the effective index, $\tilde{n}$, is defined as
\begin{equation}
\frac{1}{\langle \hs^{2} \rangle}\frac{\dd}{\dd \log R}\langle \hs^{2} \rangle=-(\tilde{n}+3).
\end{equation}
We will see in the following that this feature, the fact that reduced cumulants remain finite, 
extends to the whole generating function.

\subsection{Analytic properties of $\varphi(\lambda,\mu)$}

\begin{figure*}
\centerline{ \psfig{file=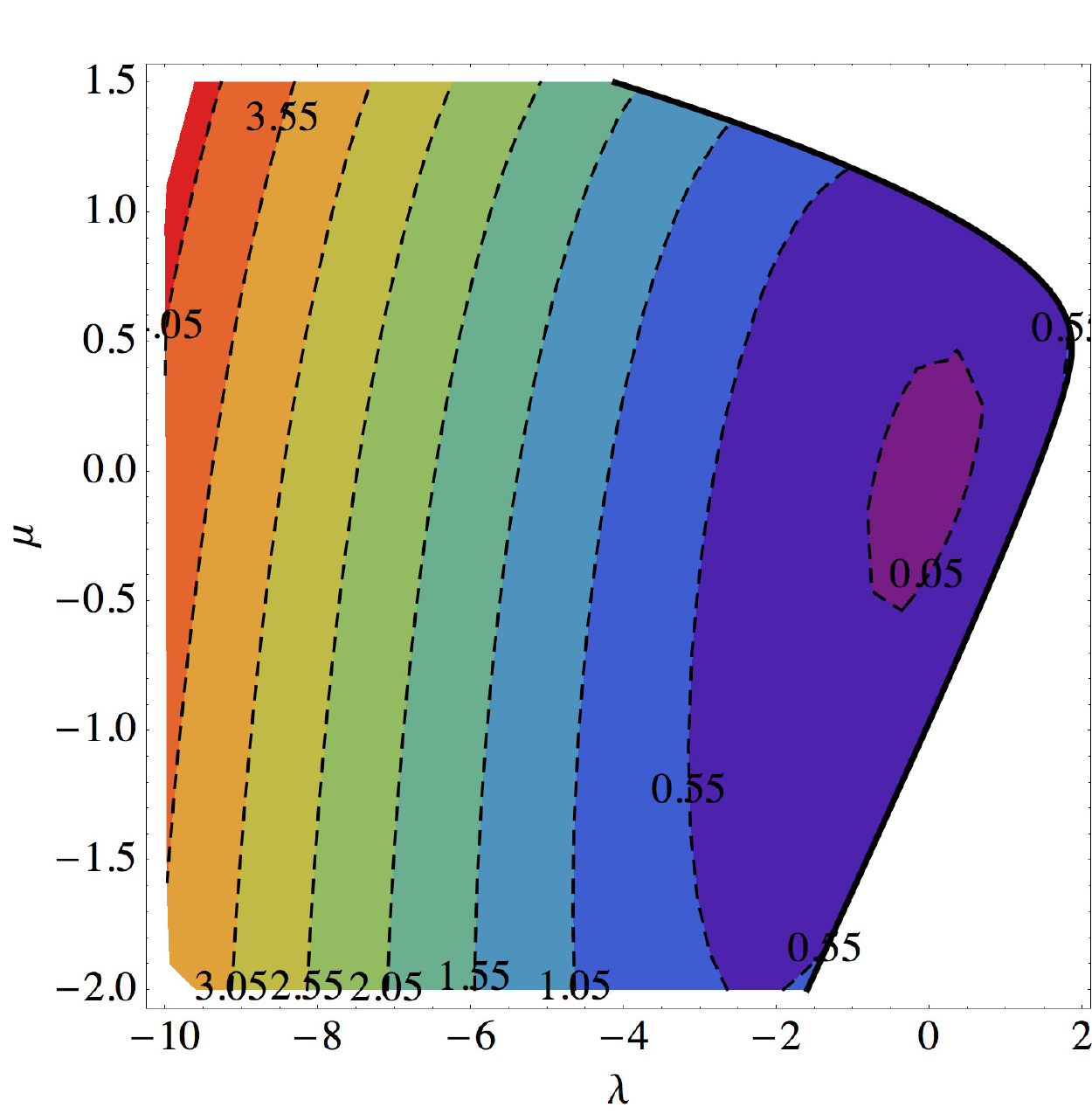, width=7 cm} \hskip 0.5cm\psfig{file=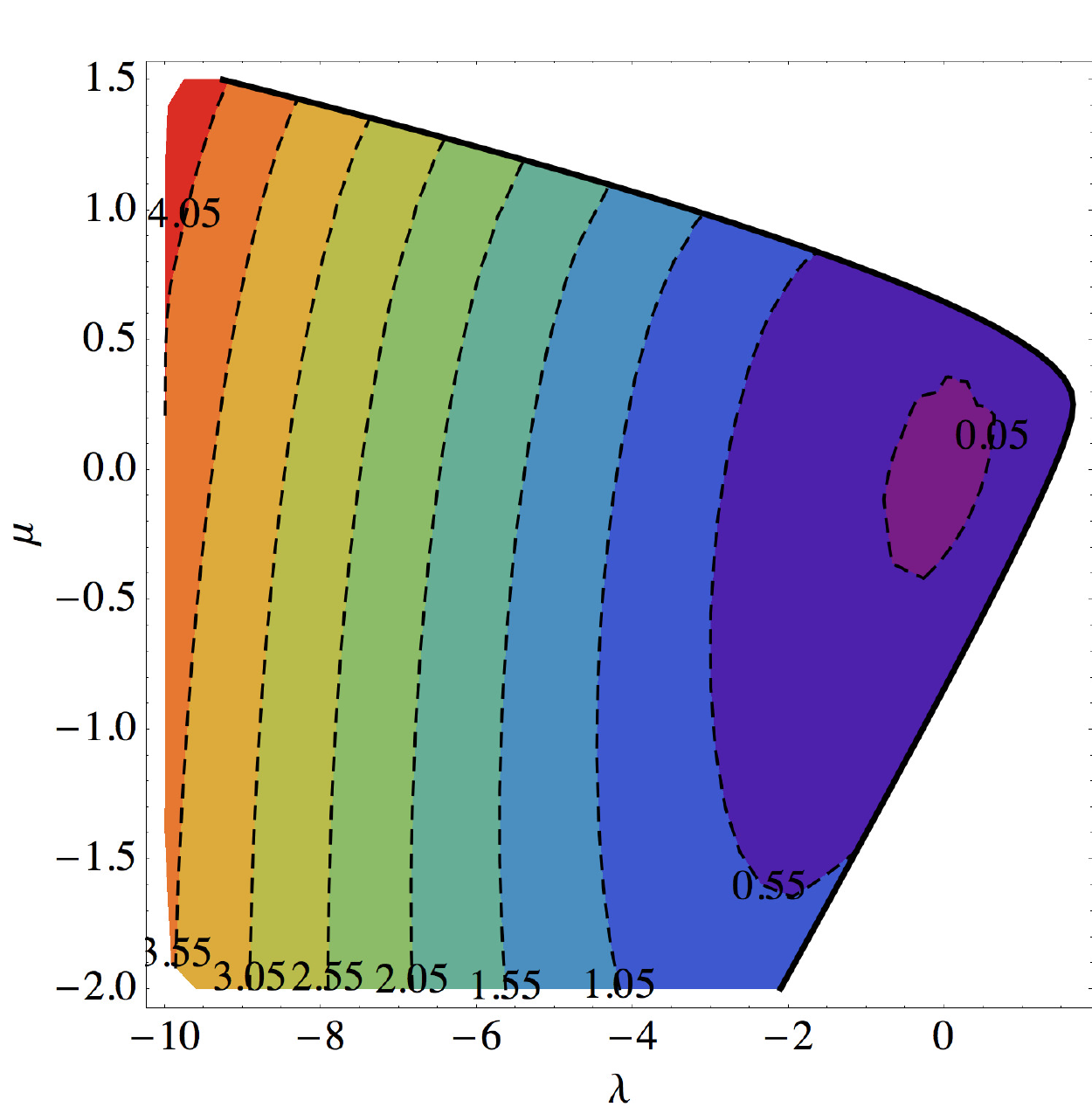, width=7 cm}}
  \caption{Contour plot of $\varphi(\lambda,\mu)-\lambda$, {\sl left} with a finite radius difference $\Dr/R=1/10$ and
{\sl  right} with $\Dr/R\to 0$. We see that the structure of the critical region, although deformed, is preserved.
 In both cases,  the restriction of $\varphi(\lambda,\mu)$ to $\mu=0$ is precisely the one-cell cumulant generating function 
  considered in Sect. \ref{Sect:OneCell}.
   \label{plotsMapPhiTheo} }
\end{figure*}

Let us now turn to the full analytical properties of  $\varphi(\lambda,\mu)$,
 for a finite radius difference to start with, and then in the limit of vanishing radius difference. 
It is to be noted that, as for the one-cell case, not all values of $\lambda$ and $\mu$ are accessible. This is due to the fact that 
the $\rho_{i}$ -- $\lambda_{i}$ relation cannot always be inverted via Eq.~(\ref{statCond3}). The boundary of
the region of interest is signaled by the fact that the determinant of the transformation vanishes,
i.e., $\det\left[{\partial^{2}}\Psi(\{\rho_{k}\})/{\partial\rho_{i}\partial\rho_{j}}\right]=0$. This condition is met  for finite values of both
$\rho_{i}$ and $\lambda_{i}$. The resulting critical line are shown as a thick solid lines on Figs. \ref{plotsMapPhiTheo}. Note that 
$\varphi(\lambda_{1},\lambda_{2})$ is also finite at this location. Within this line $\varphi$ is defined; beyond this line it is not.  
Let us now  explore the behavior of $\varphi(\lambda,\mu)$ when $\Delta R/R\to 0$. This is actually a
cumbersome limit to take. One of the reasons is that the matrix $\Xi_{ij}$ then becomes singular. 
More precisely the determinant of the cross-correlation function takes the form,
\begin{equation}
\det\left[\Sigma_{ij}(R,R+\Dr)\right]=R^{-2(3+n_{s})} \left(\frac{\Delta R}{R}\right)^{2}\frac{-9+n_{s}^{2}}{4(1+n_{s})}\,,\nonumber
\end{equation}
at leading order in $\Dr/R$ and when $n_{s}<-1$.
For a power law spectrum, the actual coefficients read
\begin{widetext}
\begin{eqnarray}
\Xi_{11}(R,\Dr)&=&\frac{2 \left(n_s+1\right) R^{n_s+3}}{{\left(\frac{\Dr}{R}\right)}^2 \left(n_s^2-9\right)} \left({\left(\frac{\Dr}{R}\right)}^2 \left(n_s^2+7 n_s+12\right)-2{\frac{\Dr}{R}} \left(n_s+3\right)+2\right)\,,\\
\Xi_{12}(R,\Dr)&=&   -\frac{R^{n_s+3}}{2{\left(\frac{\Dr}{R}\right)}^2 \left(n_s^2-9\right)} \left({\left(\frac{\Dr}{R}\right)}^2 \left(n_s^3+8 n_s^2+23 n_s+24\right)-4{\frac{\Dr}{R}} \left(n_s^2+4
   n_s+3\right)+8 \left(n_s+1\right)\right)\,,\\
\Xi_{22}(R,\Dr)&=&    \frac{4 \left(n_s+1\right) R^{n_s+3}}{{\left(\frac{\Dr}{R}\right)}^2 \left(n_s^2-9\right)}.
\end{eqnarray}
All these coefficients are diverging like $(R/\Dr)^{2}$. What we need to compute is however $\Psi(\rho,s)$ for finite values
of $\rho$ and $s$. In this case $\rho_{2}$ is also infinitely close to $\rho_{1}$  with $\rho_{2}-\rho_{1}=s {\Delta R} / {R}$
with a fixed value for $s$.
Then the resulting value of $\Psi(\rho,s)$ is finite in the limit $\Dr\to 0$. Assuming the form (\ref{zetaform}) for $\zeta(\tau)$
one gets
\begin{equation}
\begin{split}
\Psi(\rho,s)=\frac{R^{3+n_{s}} \rho ^{n_{s}/3+{(\nu -2)}/{\nu }}}
{2 \left(n_s^2-9\right) (s+3 \rho )^2}
 \left\{s^2 \left[\nu ^2 n_s^3 \left(\rho
   ^{\frac{1}{\nu }}-1\right)^2+3 n_s \left(5 \nu ^2 \left(\rho ^{\frac{1}{\nu }}-1\right)^2+16 \nu  \left(\rho
   ^{\frac{1}{\nu }}-1\right)+12\right)
   \right.\right.\\
   \left.\left.
   +4 \nu  n_s^2 \left(\rho ^{\frac{1}{\nu }}-1\right) \left(2 \nu  \left(\rho
   ^{\frac{1}{\nu }}-1\right)+3\right)+36 \left(\nu  \left(\rho ^{\frac{1}{\nu }}-1\right)+1\right)\right]+9 \nu ^2 \rho ^2
   n_s \left(n_s^2+8 n_s+15\right) \left(\rho ^{\frac{1}{\nu }}-1\right)^2
  \right.\\
   \left.   
   +6 s \nu  \rho  \left(n_s+3\right) \left(\rho
   ^{\frac{1}{\nu }}-1\right) \left(\nu  n_s^2 \left(\rho ^{\frac{1}{\nu }}-1\right)+n_s \left(5 \nu  \left(\rho
   ^{\frac{1}{\nu }}-1\right)+6\right)+6\right)\right\}.
\end{split}
\end{equation}
\end{widetext}

The function $\varphi(\lambda,\mu)$ can then be obtained by Legendre transform. {Like for the {one cell} case, the transformation becomes critical when the inversion of the stationary condition is singular. For the new variables, it is also occurring when the determinant of the second derivatives of $\psi$
vanishes
\begin{equation}
\det\left[\frac{\partial^{2}\psi(\rho,s)}{\partial \rho\partial s}\right]=0,
\end{equation}
which generalizes the condition (\ref{rhocdef}). This condition defines the location of the critical line which can then 
be visualized in the $\lambda-\mu$ plane (thick lines on Fig. \ref{plotsMapPhiTheo}). Note that the no-shell crossing
condition, which in this limit reads $s>-3\rho$, is located beyond this critical line and is therefore not relevant.}

In the regular region, the contour lines of $\varphi(\lambda,\mu)$ are shown on Fig.~\ref{plotsMapPhiTheo} 
for both a finite ratio $\Dr/R$ and when it is infinitely small. This figure explicitly  shows in particular that the limit $\Delta R\to 0$
is non pathological, in the sense that the location of the critical line and the actual value of the cumulant generating function converge
to well defined values in that limit. The convergence is however not very rapid and in practice we will use finite differences for comparisons with simulations.

\begin{figure}[ht]
\centerline{ \psfig{file=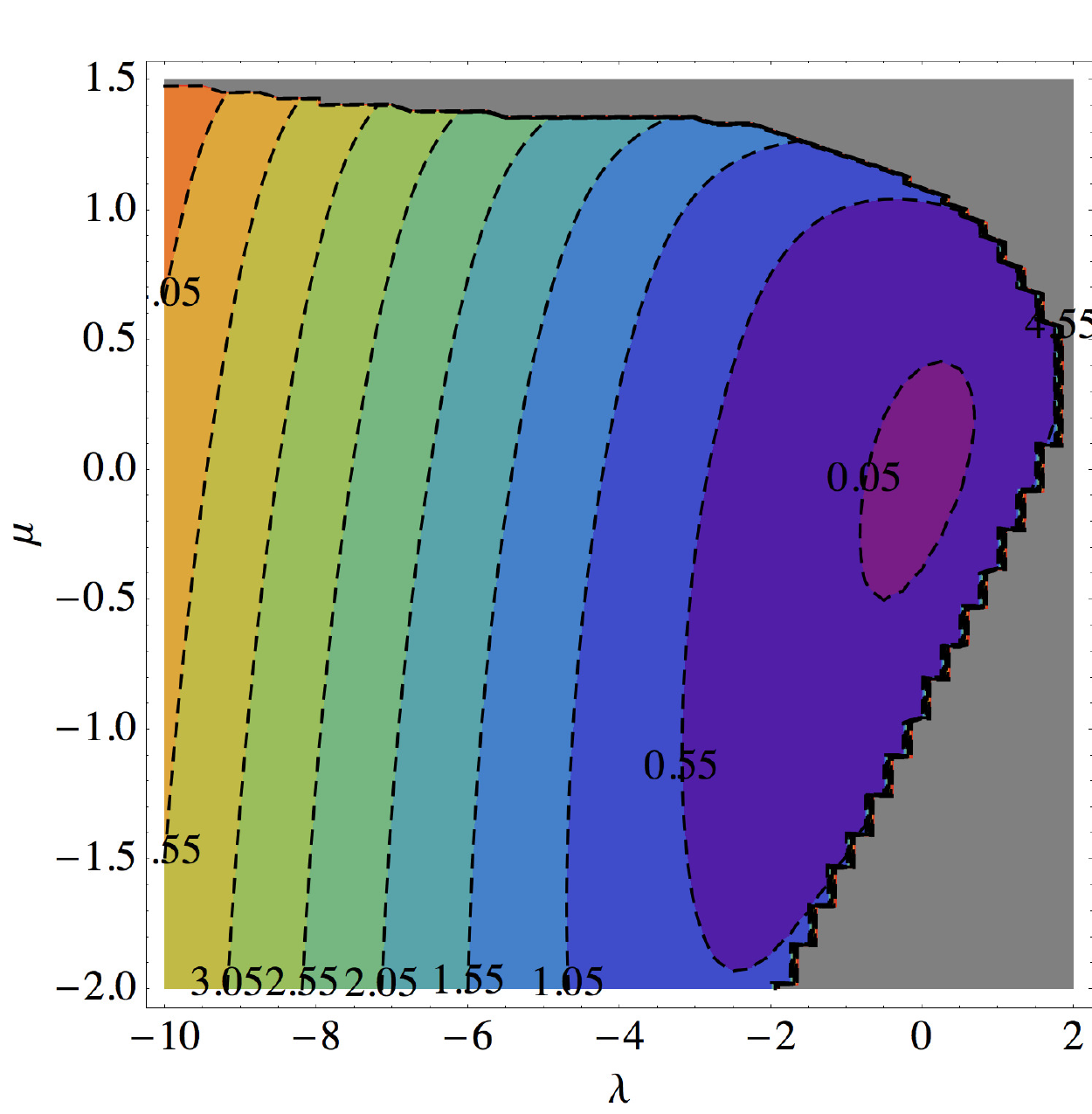, width=7 cm} }
   \caption{Contour plot of $\varphi(\lambda,\mu)-\lambda$, from the simulation. It is to be compared with left panel of Fig. \ref{plotsMapPhiTheo} 
   \label{plotsMapPhiNum}}
\end{figure}

Finally, to conclude this subsection we also compare these contour plots with  those measured in simulations. There, one actually
computes the explicit sum
\begin{equation}
\exp[\varphi(\lambda,\mu)]=\frac{1}{N_{x}}\sum_{x} \exp(\lambda\hrho_{x}+\mu \hs_{x})\,,
\end{equation}
where $\hrho_{x}$ and $\hs_{x}$ are the measured values of $\hrho$ and $\hs$ in a cell centered on $x$ (in practice on grid points)
and $N_{x}$ is the number of points used (see Appendix~\ref{Simulations} for details). 
Then $\varphi(\lambda,\mu)$ is always well defined, irrespectively of the values of $\lambda$ and $\mu$. To detect the location of a critical line
one should then rely on the properties it is associated to. From the analysis of the one cell case it appears  that for $\lambda>\lambda_{c}$,
$\varphi(\lambda)$ is ill defined because $\int\mP(\hrho)\exp(\lambda\hrho)\,\dd\hrho$ diverges. More precisely when $\lambda
\to\lambda_{c}$
the value of $\varphi(\lambda)$ becomes dominated by the rare event tail. It makes such a quantity very sensitive to  cosmic variance and
in practice the critical line position is therefore associated with a diverging cosmic variance. 
In the two-cell case,  we encounter the same effects. To locate we therefore simply cut out part of the $(\lambda-\mu)$ plane for which 
the measured  variance of $\varphi(\lambda,\mu)$
exceeds a significant fraction of its measured value. We set this fraction to be 20\%~\footnote{the location of the resulting critical line is only weakly sensitive 
to the threshold we choose.}. This criterium give rises to the solid line shown on Fig. \ref{plotsMapPhiNum}.
This figure is now to be compared to the left panel of Fig. \ref{plotsMapPhiTheo}. Although the figures are not identical they clearly exhibit the same patterns.

\subsection{Slope cumulant generating function and PDF}

\begin{figure}[ht]
\centerline{ \psfig{file=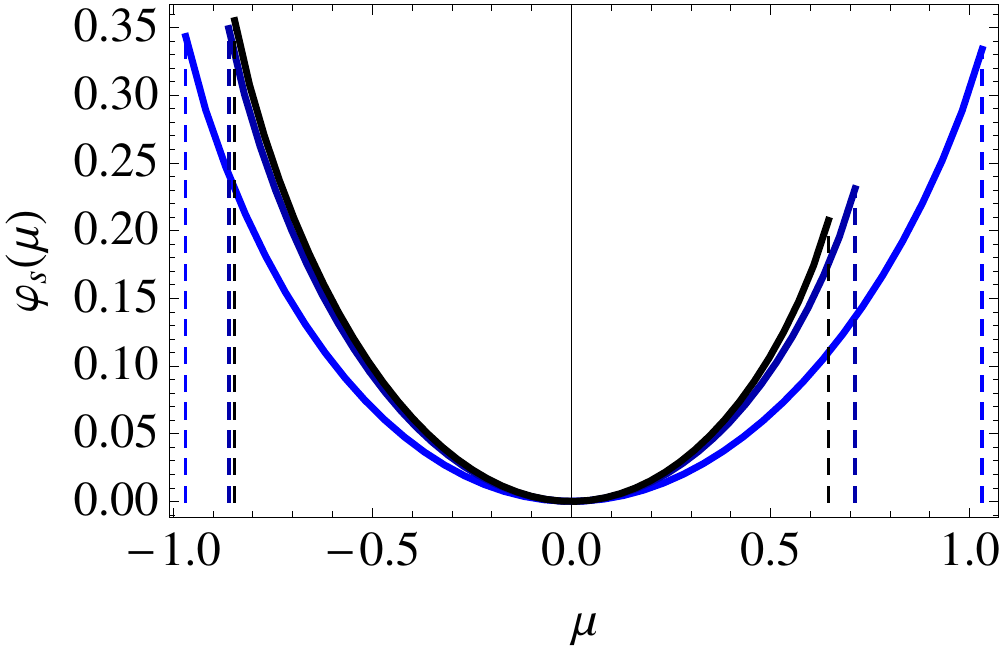, width=8 cm} }
   \caption{The slope generating function $\varphi_{s}(\mu)$ for finite differences, $\Delta R/R=0.1$ and $\Delta R/R=0.01$,
   and in the limit $\Delta R/R\to 0$. The corresponding curves are respectively in blue, darker blue and black. The vertical dashed lines
   show the locations of the critical points, $\mu_{c}^{-}$ and $\mu_{c}^{+}$.}
   \label{phidemuLimit}
\end{figure}

 When one wishes to build the PDF
of $\hs$, one needs to restrict $\varphi(\lambda,\mu)$   presented in the previous section  to the $\lambda=0$ axis, i.e.  focus on
$\varphi_{s}(\mu)\equiv\varphi(\lambda=0,\mu)$.  Fig.~\ref{plotsMapPhiNum}  shows that $\varphi_{s}(\mu)$
has two extrema points, one corresponding to a positive value of $\mu$, $\mu_{c}^{+}$, and one to a negative value $\mu_{c}^{-}$. 
The resulting global shape of $\varphi_{s}(\mu)$ is shown 
on Fig. \ref{phidemuLimit}, where it is also compared to the results where $\Delta R/R$ is kept
finite. It actually shows that  the limit $\Delta R/R$ is genuine at the level of the cumulant generating function
but is reached for very small values of $\Dr/R$. When predictions are compared with simulations for which
the slope is measured with finite differences, it is necessary to use a finite difference $\Dr$. 

\begin{figure}[ht]
\centerline{ \psfig{file=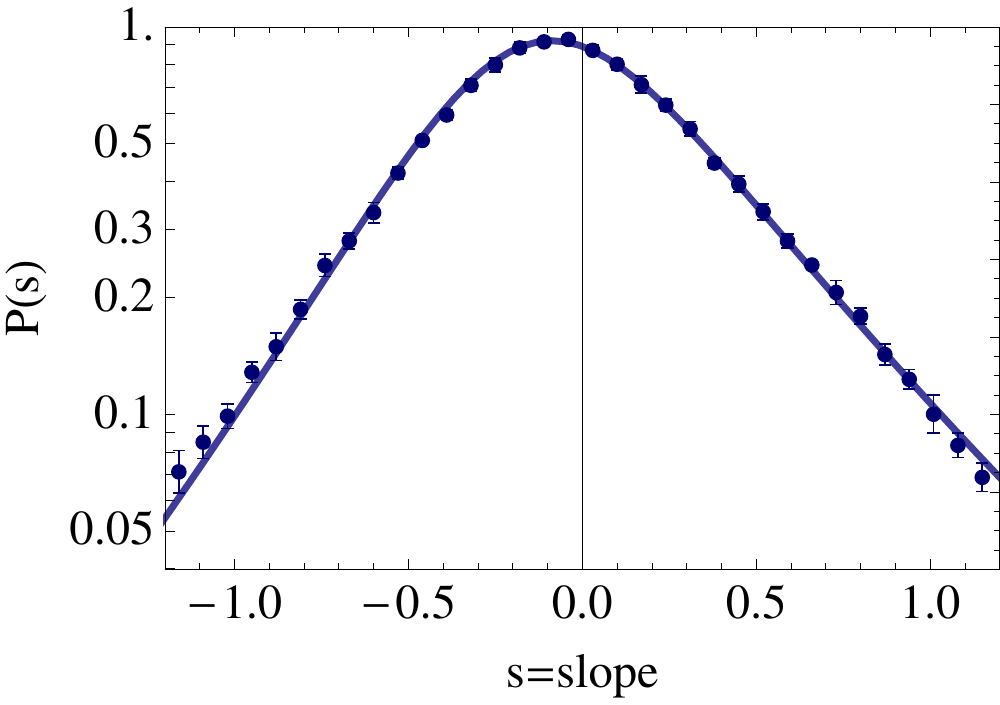, width=7 cm}}
\centerline{ \psfig{file=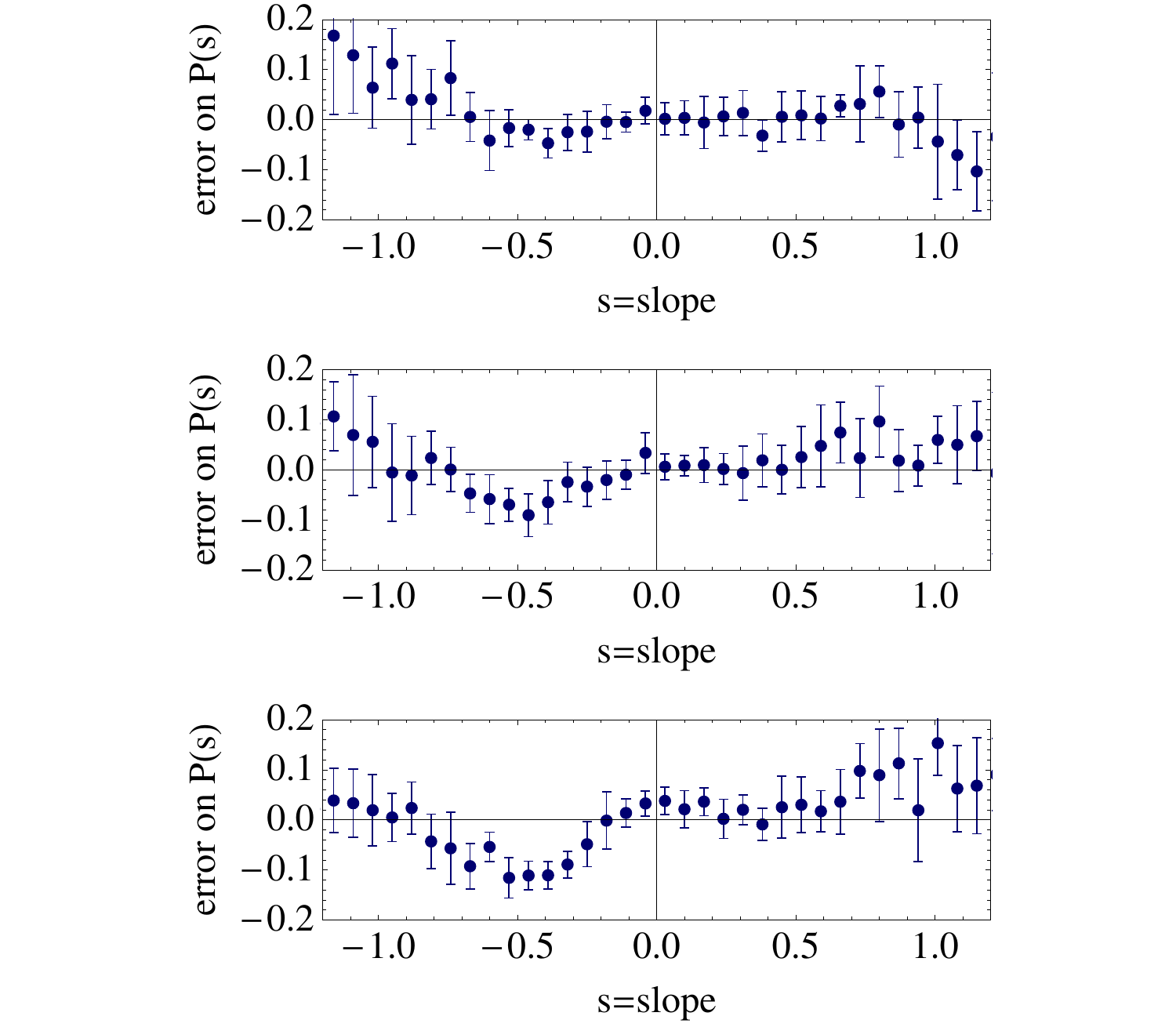, width=9 cm}}
  \caption{The PDF of the slope  for $z=1.46$. The bottom panels show the residuals for $z=1.46$, $z=0.97$ and $z=0.65$ from top to bottom.  
   \label{PDFSlope}}
\end{figure}

We are now in position to build  the one-point PDF of the density {\em slope} via the inverse Laplace  transform of the cumulant generating function.
It should be clear from the singular behavior of $\varphi_s(\mu)$ that it will exhibit exponential cut-offs on both sides, for 
positive and negative values of $\hs$ although not a priori in a symmetric way.  
In practice, to do the complex plane integration, 
we build the function $\varphi_{s}(\mu)$ for the actual power spectrum of interest,
 and {\sl then}  build an effective form $\eff{\zeta}(\tau)$
that reproduces the numerical integration following Eqs.~(\ref{hzetaform1})-(\ref{hzetaform2}) as explained in \cite{2000A&A...364....1B}. 
In practice we use a 7th order polynomial to do the fit. We then proceed via  integration in the complex plane 
using the usual approach {(see Appendix~\ref{Integrationinthecomplexplane})}. The results for  $R=10 h^{-1}$ Mpc and $z=1.46$ and $z=0.97$ is presented on the top panel of Fig. \ref{PDFSlope}.
The figure clearly exhibits the expected double cut-offs. Discrepancies between numerical results and theory that can be seen 
in the bottom panels for $\hs\approx -0.5$ are not clearly understood (cosmic variance, numerical artifacts?).

\subsection{The expected constrained slope and profile}
\label{sec:ConsProf}

\begin{figure}[ht]
\centerline{ \psfig{file=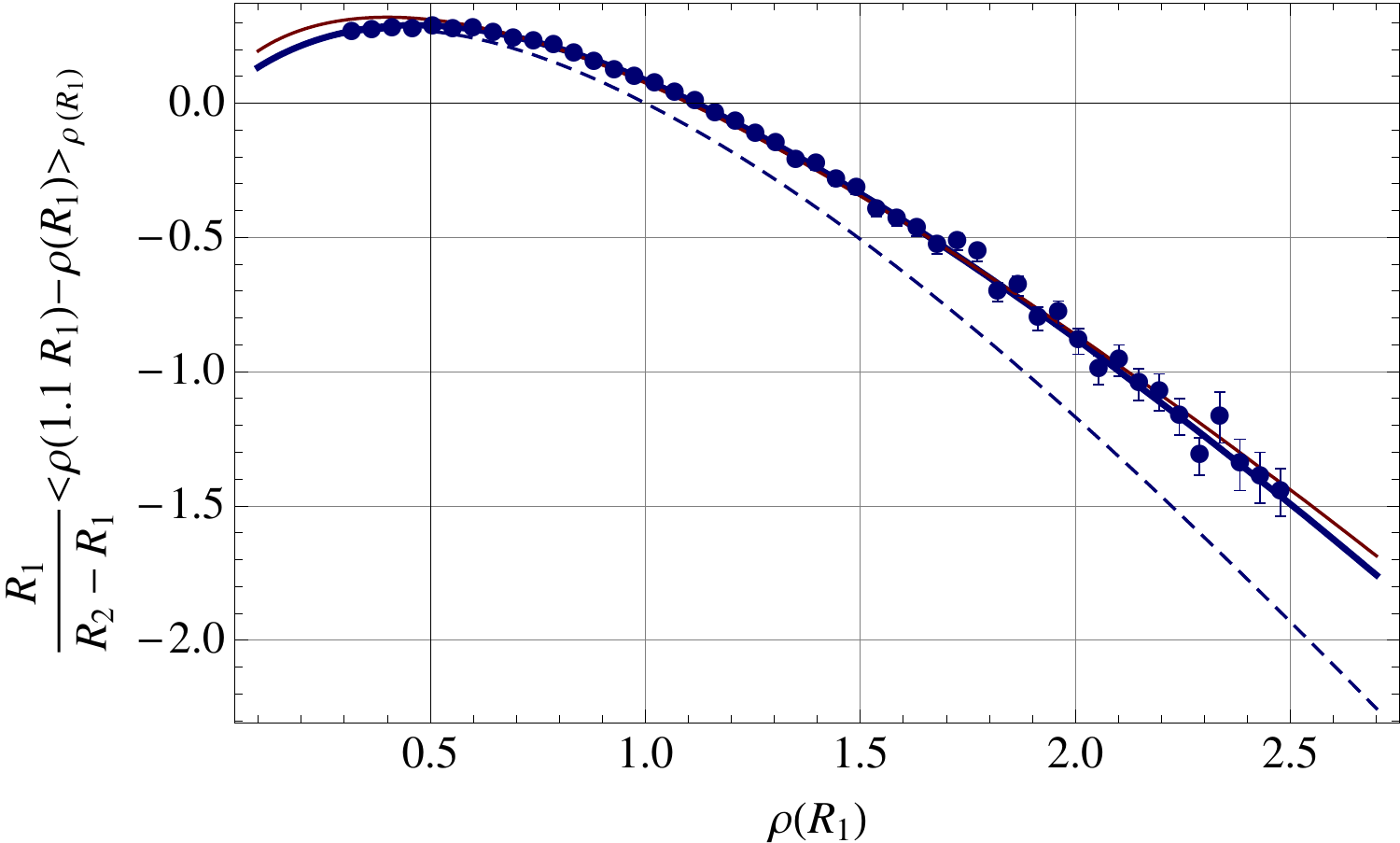, width=7.5 cm} }
\centerline{ \psfig{file=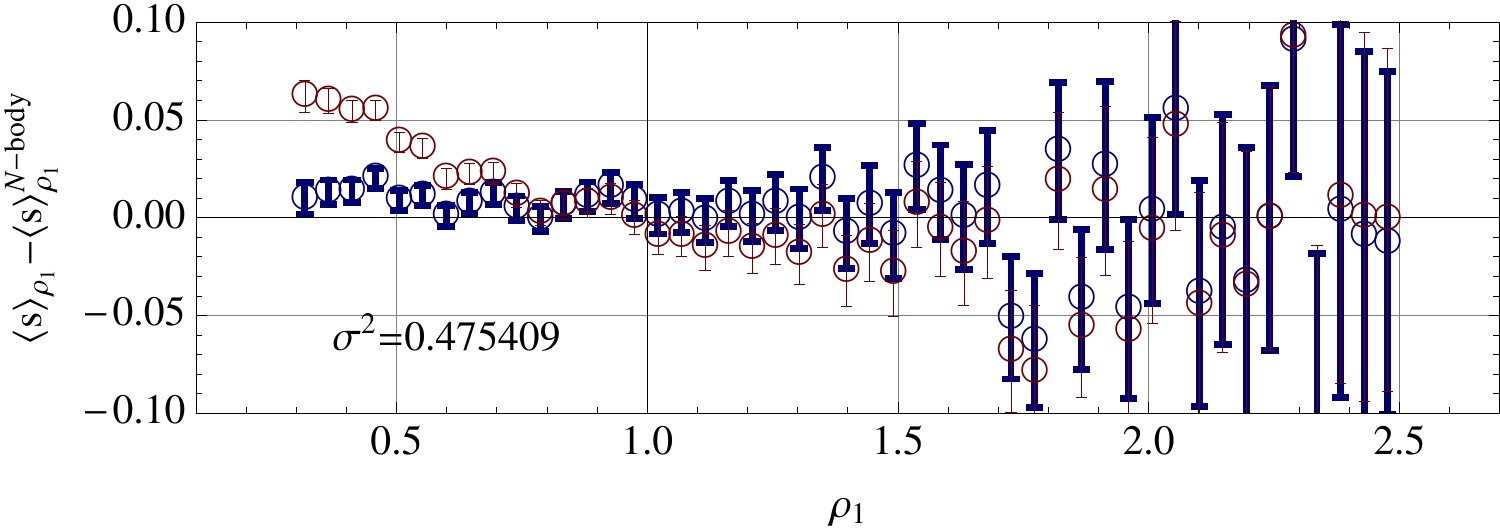, width=7.5 cm} }
\centerline{ \psfig{file=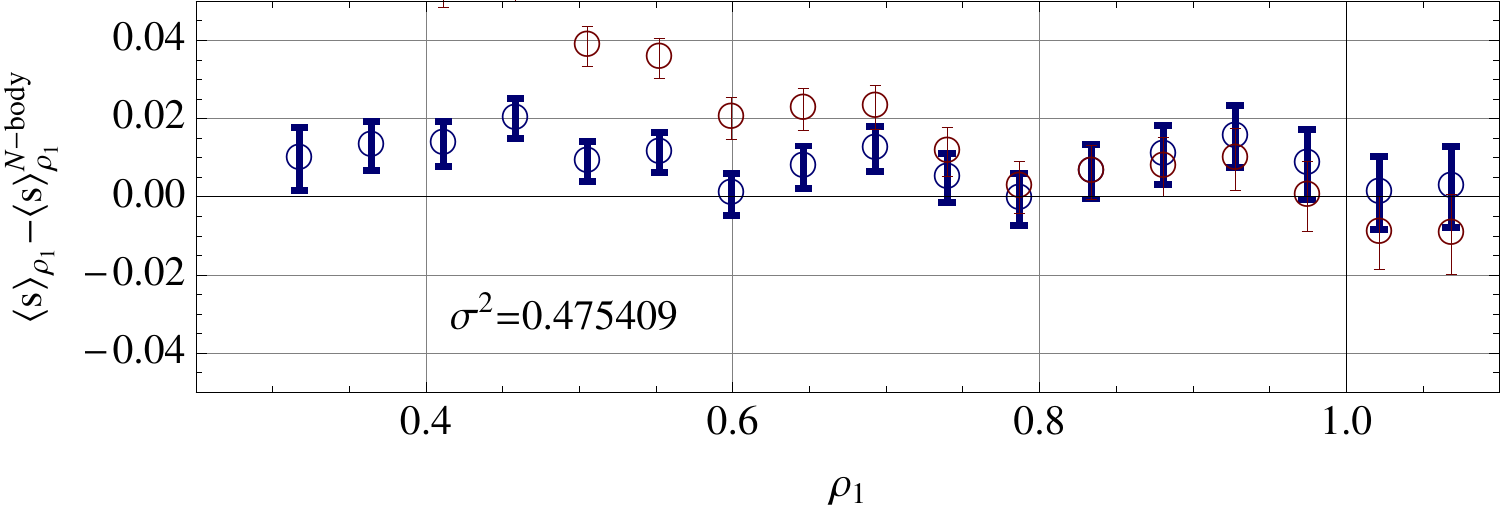, width=7.5 cm} }
   \caption{Top: the conditional profile, $\langle \hs\rangle_{\hrho(<R_{1})}$ as a function of $\hrho(<R_{1})$. The thick blue solid line  is the result of the numerical integration; the thin dashed line the saddle point approximation Eq. (\ref{saddlemeanprof}). We also present the power law approximation case as a thin (red) solid line. It is shown to depart from the exact prediction in the low density region.
    {the agreement between the theory and the measurements  near the origin is quite remarquable.}
{The bottom panels} show the residuals computed in bins 
   as a function of the density (with a zoomed plot below). Again the thick symbols are correspond to the exact calculation, the thin symbols correspond to the power low approximation.
    \label{ExpProfil1}}
\end{figure}

\begin{figure}[ht]
\centerline{\psfig{file=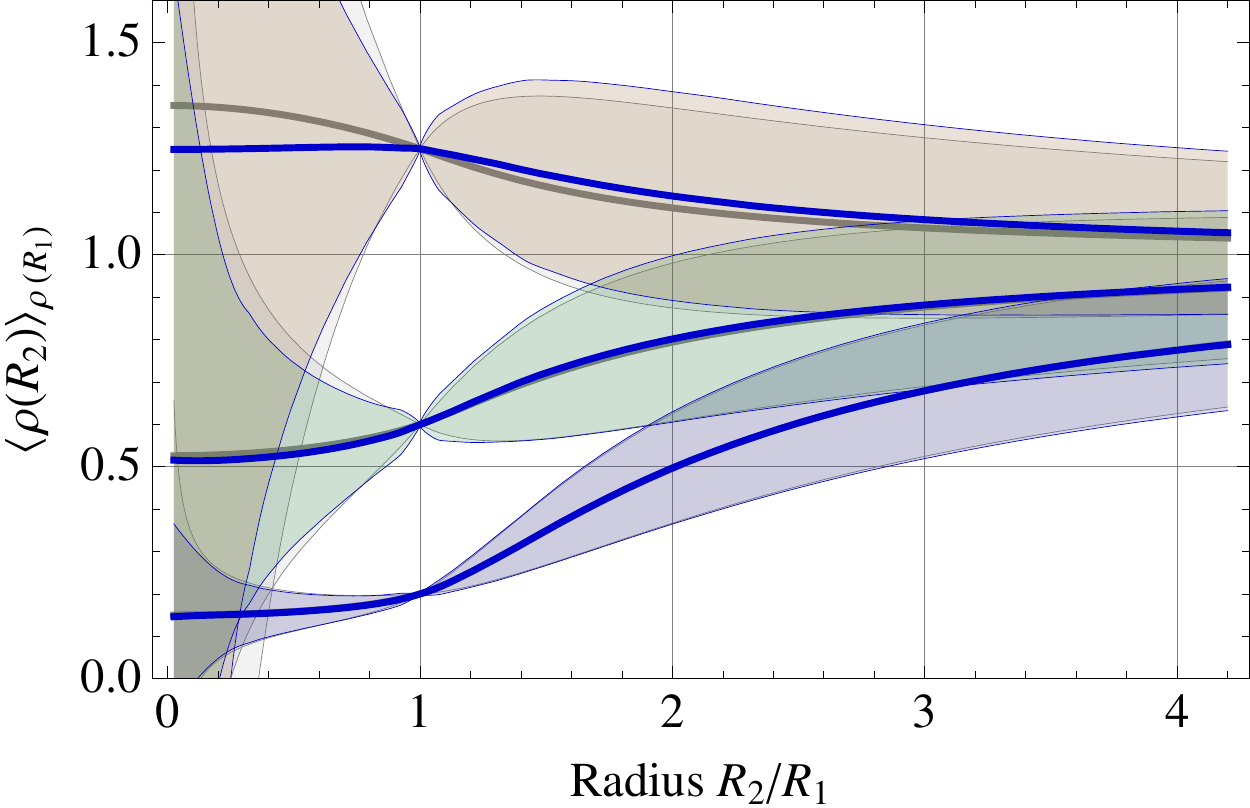, width=8.5 cm}}
\caption{The conditional profile as a function of $R_{2}$ and for different choices of $\hrho(R_{1})$ (to which $\hrho(R_{2})$ is equal to at $R_{2}=R_{1}$). The blue thick solid lines are the results of numerical integrations; the colored thin lines show the 1-$\sigma$ variance about the expectation. The close-by gray lines are the same calculations but using the saddle point approximations of  Eqs. (\ref{saddlemeanprof}) and (\ref{saddlevarprof}) respectively. Note in particular the smaller variance of the underdense 
profile near $R_2\sim 0$.     \label{ExpectedProfils}}
\end{figure}

\begin{figure}[ht]
\centerline{ \psfig{file=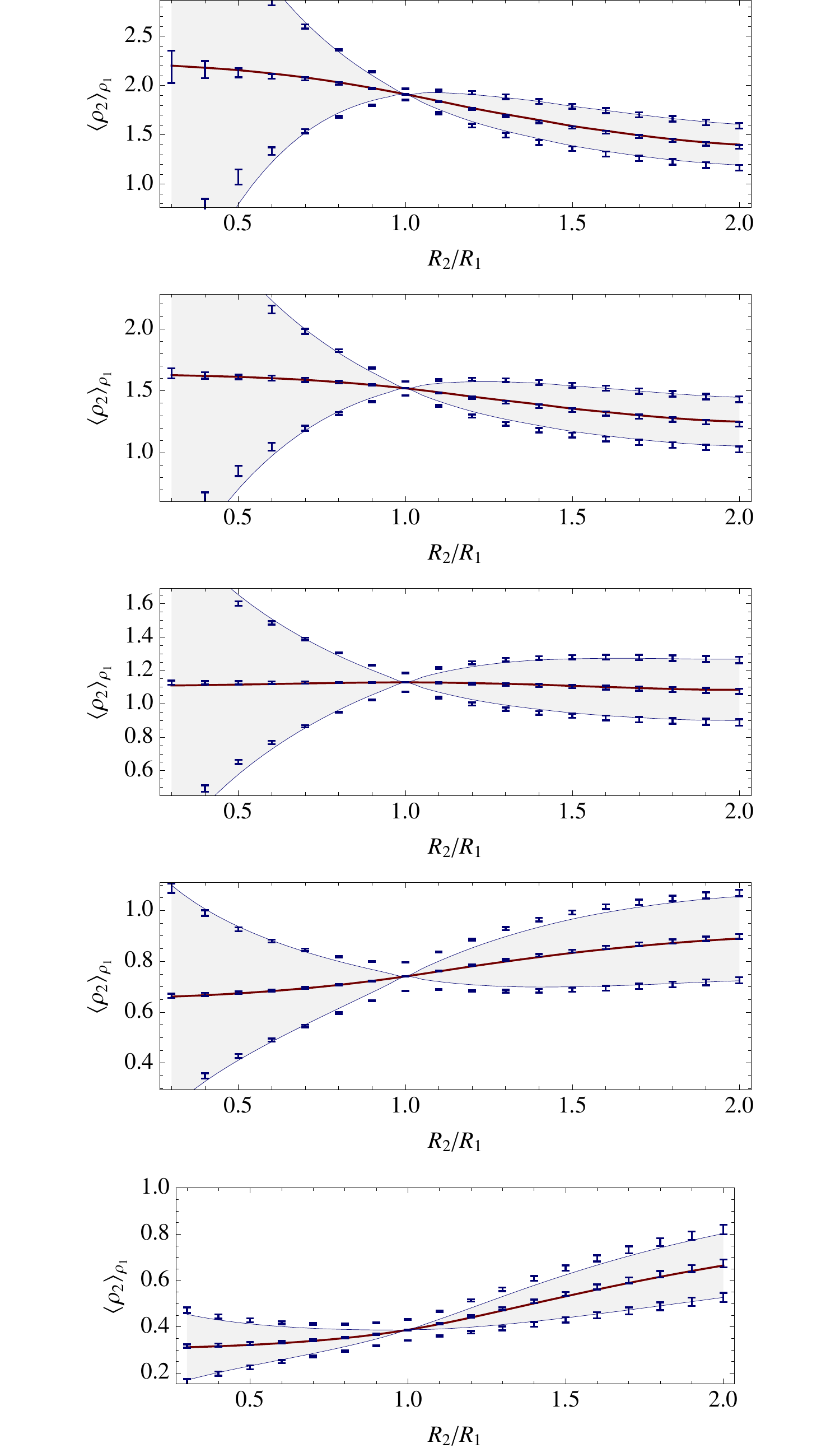, width=8.5 cm} }
   \caption{Same quantities as in the  Fig.~\ref{ExpectedProfils} measured here in simulations. The solid lines are the theoretical predictions and the points with error bars are the measurements for both the expected  value and its variance. The agreement is spectacular in particular for low density constraints.
    \label{ProfilesTheoNum}}
\end{figure}

Let us finally move to the key result of this paper. In the previous subsection we built the marginal PDF of $\hs$; 
we now focus on the conditional properties of $\hs$ given $\hrho_{1}=\hrho(<R_{1})$ at a given $R=R_{1}$, whether $\hs$ is defined from a nearby radius of not. 
Mathematically it can be expressed in terms of the joint PDF, $\mP(\hrho_{1},\hrho_{2})$, as
\begin{equation}
\langle \hs\rangle_{\hrho_{1}}=-\frac{R}{\Dr}\hrho_{1}+\frac{R}{\Dr\, \mP(\hrho_{1})}\int\dd \hrho_{2} \ \hrho_{2} \ \mP(\hrho_{1},\hrho_{2})\,,
\end{equation}
given that 
\begin{eqnarray} \hskip -0.5cm
\int\dd \hrho_{2} \ \hrho_{2} \ \mP(\hrho_{1},\hrho_{2})&=& \nonumber\\ && \hskip -3.75cm
\int_{-\ii\infty}^{+\ii\infty}\frac{\dd\lambda_{1}}{2\pi \ii}
\left.\frac{\partial \varphi(\lambda_{1},\lambda_{2})}{\partial\lambda_{2}}\right\vert_{\lambda_{2}=0}
\exp(-\lambda_{1}\hrho_{1}+\varphi(\lambda_{1}))\,, \label{defcondrho2}
\end{eqnarray}
which can be obtained by explicit integration in the complex plane \footnote{it involves expressing $P$ as a function of $\varphi$,
expressing $\rho_2$
 integrating over $\rho_2$, and using Cauchy's theorem}. Note that the solution of the stationary equations, Eq.~(\ref{statCond3bis}), yields  the identity
\begin{equation}
\left.\frac{\partial \varphi(\lambda_{1},\lambda_{2})}{\partial\lambda_{2}}\right\vert_{\lambda_{2}=0}=
\rho_{2}(\lambda_{1},\lambda_{2}=0)\,. \label{eq1storder}
\end{equation}
 For the saddle point solution corresponding to the low $\rho$
regime, $\lambda_{1}$ and $\hat \rho_{1}$ in Eq.~(\ref{defcondrho2}) are related through the stationary condition. In this limit we therefore have
\begin{equation}
\langle \hrho_{2}\rangle_{\hrho_{1}}=\overline\rho_{2}(\hrho_{1})\,,
\label{saddlemeanprof}
\end{equation}
where $\overline\rho_{2}(\hrho_{1})$ is the solution of the system
\begin{equation}
\lambda_{1}=\frac{\partial\Psi(\rho_{1},\overline\rho_{2})}{\partial\rho_{1}},
\ \ \ 0=\frac{\partial\Psi(\rho_{1},\overline\rho_{2})}{\partial\rho_{2}}.
\end{equation}

These calculations can be extended to the constrained variance of the slope. The computation follows the same line
of derivation but is slightly more involved. It is presented in appendix \ref{Ap:ConsVar}.

Let us now present the expected slope from exact complex plane numerical integration, using  the analytical saddle point 
approximations and as measured in numerical simulations. For instance,  Fig.~\ref{ExpProfil1}  shows the expected slope  
given by $10. \times [\hrho(1.1 R)-\hrho(R)]$ as a function of $\hrho(R)$ using the same cosmological parameters as for Fig.~\ref{Prho1D}. 
The solid lines are the results of complex plane integrations and the dashed line is the saddle point approximation. The latter is found to perform very poorly when compared to simulation.  We note also  that the low density part of the prediction can only be accounted for when the running parameter is taken
into account. This is clearly visible in the middle and bottom panels when one compares the (thick) blue and the (thin) red marks.  These
comparisons show that the analytical predictions are accurate at percent level in a large range of parameters.

Let us finally turn to a more global properties of the density in cells and consider the density \emph{profile} defined
as the constrained density $\hrho(R_{2})$ given $\hrho(R_{1})$ as a function of $R_{2}$. Technically computing
expected profiles or slopes is equivalent. The second point of view allows however to visualize what should be the radial variation of the
density profile, and its fluctuations, of an under dense or an over dense region.
The result of such a calculation is presented on
Fig.~\ref{ExpectedProfils}  which shows the expected density as a function of the radius $R_{2}$ and for various values
of $\hrho(R_{1})$. In the same plot we also show the expected 1-$\sigma$ variance about the expectation values. Both quantities are computed using the exact complex plane integration and compared to their saddle point approximation counterparts. The difference is only significant
for $\hrho(R_{1})=1.25$.
Interestingly for low density prior, e.g. $\hrho(R_{1})=0.25$ on the figure, the variance is small (and significantly smaller than the variance
of $\hs$ in the absence of prior on the density). That implies that all voids should look similar, probably a good starting
point for exploring the statistical properties of the field while focussing on these regions.

Comparisons of the latter prediction 
with numerical simulations is made in Fig.~\ref{ProfilesTheoNum} where we give both the measurements of the  
expected profile and their variance for a given constraint. The only difference with the theoretical predictions is that the constraints is binned, i.e. the prior is that the density $\rho_{1}$ is assigned in a given bin of width $0.2$ centered on the values $0.35, 0.74, 1.13, 1.52, 1.92$. As the theoretical predictions do not take into account the binning, there is a noticeable departure between the predicted variance and its measured value near $R_{2}/R_{1}\approx 1$ due to the width of the bin. But,  this departure notwithstanding, the agreement between the theoretical predictions and the measured quantities, for
both the expected profile and its variance is just striking! Only when the constraint density is large (top 2 panels),  can we see some slight
departure with the theory for small radii, which  is due to the fact that they correspond to regions  entering the nonlinear regime.

\subsection{Joint $n$-cells PDF}

The results presented in the previous section give us confidence in the general framework we have adopted
here.  It has to be stressed that all of the
properties we have described are simultaneously captured with the shape of the multiple cell cumulant generating function,
$\varphi(\{\lambda_{k}\})$ or its counterpart, the $n$-cell PDF, $\mP(\{\hrho_{k}\})$.  We should keep in mind that we could have considered
the former only to compare with simulations but we dramatically lack intuition for such a representation. By contrast we have much better intuition of what $n$-cells PDFs are. So far we have considered only the one-cell PDF.
In the following we succinctly consider the derivation of the multi-cell PDF in our framework.

Hence  let us consider a set of $n$ concentric cells and its cumulant generating function $\varphi(\{\lambda_{k}\})$. 
In principle the corresponding PDF, $\mP(\{\hrho_{k}\})$,  
is to be obtained from inverse Laplace transform. Such a computation appears extremely
challenging to implement and we have not succeeded yet in producing a full 2-cell PDF. 
We can however present its low density approximation, the counterpart of Eq.~(\ref{aPDFlowrho}), for a multi-dimensional case. It is based
on the use of the saddle point approximation of  Eq.~(\ref{InvLapTransnD}) assuming the overall  variance is small. 
It leads  to a similar condition that should be met at the saddle point $\{\lambda_{s}\}_i$
\begin{equation}
\frac{\partial}{\partial \lambda_k}\left[\sum_{i}\lambda_{i}\hrho_{i}-\varphi(\{\lambda_{i}\})\right]=0\,,
\end{equation}
which leads to 
\begin{equation}
\hrho_{i}=\rho_{i}(\{\lambda_{k}\})\,,
\end{equation}
and with the constraint that 
\begin{equation}
\displaystyle{\rm det}\!\left[\frac{\partial^{2}\Psi}{\partial\rho_k \partial\rho_l}
\right]>0
\end{equation}
at the saddle point position.
The resulting expression for the density PDF generalizes Eq.~(\ref{aPDFlowrho}) to
\begin{equation}
\mP(\{\hrho_{k}\})\!\!=\!\!\frac{1}{(2 \pi)^{{n}/{2}}} {\sqrt{\displaystyle{\rm det}\!\left[\frac{\partial^{2}\Psi}{\partial\hat\rho_k \partial \hat \rho_l}
\right]}}
\exp\left[-\Psi(\{\hrho_{k}\})\right].  \label{ndpdfsaddle} 
\end{equation}
This analytic expression is expected to be an approximate form for the exact PDF in underdense regions.

We suggest that in the absence of computable multiple-cell PDFs, this form could be used, provided one makes sure to restrict its application 
to its proper region of validity. It is interesting to note that, in this framework,  the \emph{parameter dependence } of the mode of that PDF and its local curvature tensor can be straightforwardly computed from it analytically. In the concluding section we will simply sketch a way to constraint key cosmological parameters using this form.

\section{Conclusions and prospects}
\subsection{Summary}
In the context of upcoming large wide field surveys 
we revisited the derivation of the cumulant generating functions of densities in spherical concentric cells in the 
limit of a vanishing variance and we conjectured that  it  correctly represents the generating 
function \emph{for finite values of the variance}.  We noted that such a quantity is an observable in itself and could
probably be used as a cosmological indicator. In this study we however focused our efforts on its counterpart, the multi-cell
density probability distribution function (PDF). 

We first computed the resulting one-cell density PDF. These results were tested 
with unprecedented accuracy, in particular taking into account  the scale variation of the 
power spectrum index. Comparisons to modern $N$-body simulations showed that predictions reach percent order
accuracy (when the density variance is measured from simulations) 
for a large range of density values, as long as the variance is small enough. It confirmed in particular that this 
formalism gives a good account of the rare event tails: predictions are in agreement with the numerical measurements
down to numerical precision.

We took advantage of the finite variance generating function formalism to explore its implications  to the two-cell case
in a novel regime. In particular we derived the statistical properties
of the local density slope, defined as the infinitesimal difference of the density in two concentric cells of
 (possibly infinitesimally) close radii. We gave its mean expectation, and its expectation constrained to  a given density. 
From the properties of the local slope, one can also construct the overall expected profile, i.e. the density as a function
of the radius, and its fluctuations.
We found the latter to be of particular interest when focussing on voids, as
in these regions, fluctuations around the mean profile are significantly reduced.  In particular we suggest
below a possible method to constrain cosmological and gravity models from these low density regions.
All these predictions were successfully compared to simulations.

\begin{figure}
\centerline{\psfig{file=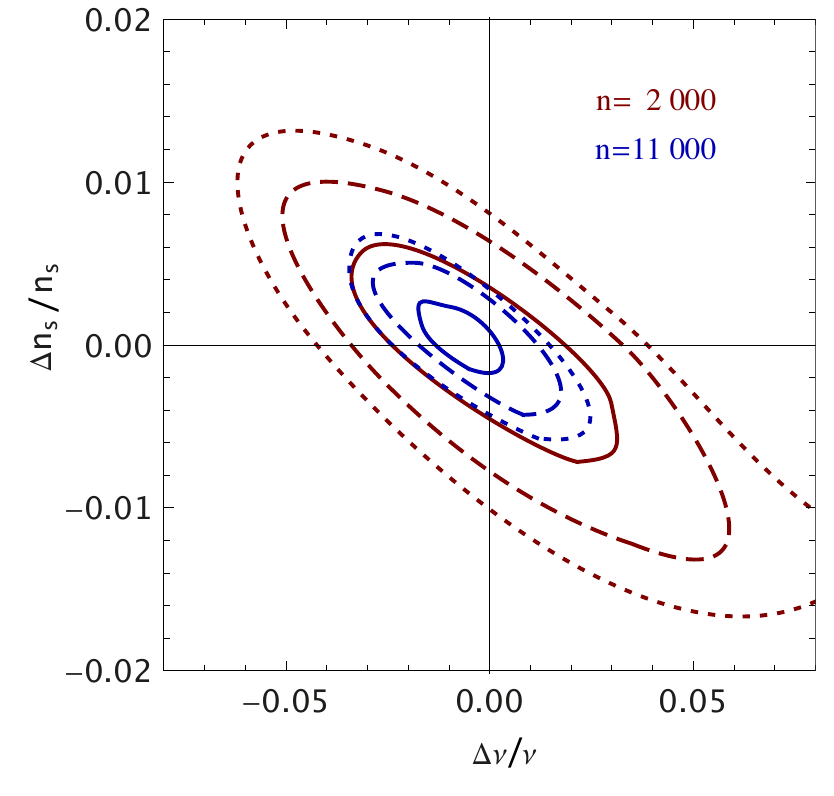, width=8 cm}}
   \caption{
      The likelihood contours at one, three and five sigmas around the reference model, $(n_s=-2.5, \nu=3/2)$ drawn from  $\sim$11\,000 measurements ({\sl inner blue contours}) and $\sim$2\,000 measurements ({\sl outer red contours})  of the 
    densities in concentric shells of radii 10 and 11 Mpc/$h$.
  }
    \label{maxlike}
\end{figure}

\subsection{Prospects}
The full statistical power of  the  approach presented in this paper would ultimately be encoded in the shape of the 2-cell 
density PDF but we do not know at this stage how to properly invert the exact expression given by Eq.~(\ref{InvLapTransnD}) in this 2-cell regime.
Despite this limitation,  as we do not have simulations that span different gravity models, let us use the saddle point form of Eq.~(\ref{ndpdfsaddle}) assuming it is exact (hence avoiding the issue of the domain of validity of that analytic fit to the exact PDF), 
and use its dependence on key (cosmic) parameters to infer the precision with which cosmological parameters could be constrained. 

Focusing the analysis on two quantities, the parameter $\nu$ that encodes the spherical collapse dynamics (see Eq.~(\ref{zetaform})) and the
power law index $n_{s}$, let us  simply consider  sets of about 2\,000 and 11\,000 independent 
measurements drawn in concentric spheres  of radii  10 and 11 Mpc/$h$ and such that $0.05<\rho_{1}<0.5$ and $-0.02<\Delta\rho/\rho_{1}<0.06$ (i.e. near the peak of the PDF).
  The sample are drawn directly from the  two-cell PDF for chosen values of the  power law power spectrum and $\nu$ parameter 
staring respectively with $n_s=-2.5$ and $\nu=3/2$. The likelihood of the models where $n_s$ and $\nu$ vary  in the range $[-0.12,0.12]$ around the reference value is computed. 

The resulting mean (over 25 independent samples)
log-likelihood of the data set as a function of $\Delta \nu/\nu$ and $\Delta n_s/n_{s}$ is displayed on
Fig.~\ref{maxlike}
(at one, three and five sigma resp.)
\footnote{a similar experiment for three-cells of size $R=11,12$ and $13$Mpc$/h$ was carried out, producing similar results.}.
 As expected, the likelihood contours are centered on the zero offset values; they yield the precision that could be reached in 
 a survey of useful volume of about $(200\, h^{-1}{\rm Mpc})^{3}$ (red contours) and  $(360\, h^{-1}{\rm Mpc})^{3}$ (blue contours). 
 These sample size are not unreasonable. 
 Indeed, the volume span with about 30\,000 spheres correspond to the volume covered by the simulation and we found that, in doing so,  the error bars on relevant quantities such as profiles (as shown on Fig.~\ref{ExpProfil1}) were of the same order as the one measured in the simulation we used throughout the
paper. At face value, relative accuracies below the percent on $(n_s,\nu)$ could be reached with such surveys.
Yet, this numerical experiment is, at this stage, at best illustrative. We are indeed aware that 
in more realistic situations, one would have to properly account for the domain of validity of 
the above functional form, which would take us beyond the scope of this paper.
Another open question would be to estimate how many concentric cells should be used to get an optimal constraint 
for a given set of  cosmic parameters but the answer to this question will probably depend on the geometry of the available  survey.

Should these problems be alleviated, effective implementation of such cosmological tests would still be far fetched. In particular 
galaxy catalogues in $z$-space break the local spherical symmetry in a complex way making the application of such method impractical.
One way to avoid this problem is to stick to observations for which this method is applicable, such as {\sl projected} densities along the line of sight. It can be done either
in the context of cosmic shear observations or for photometric like redshift surveys. In both cases the point is not to reconstruct the
spherical 3D statistics but the circular 2D statistics for which the whole method should be applicable following early investigations in~\cite{2000A&A...364....1B,1995A&A...301..309B}.
The accuracy of the predictions have still to be assessed  in this context. 
Another missing piece that can be incorporated is the large distance correlation of statistical indicators such as profiles and constrained profiles. Following  \cite{1995A&A...301..309B} it is indeed within reach of this formalism to compute such quantities. We would then have a fully working theory that could be exploited in real data sets.

{\bf Acknowledgements:} 
We warmly thank D. Pogosyan for triggering our interest in studying the statistics of void regions. We also thank him for many comments.
 This work is partially supported by the grants ANR-12-BS05-0002 and  ANR-13-BS05-0005 of the French {\sl Agence Nationale de la Recherche}
and by the National Science Foundation under Grant No. NSF PHY11-25915. The simulations were run on 
the {\tt Horizon} cluster. CP thanks KITP and the University of Cambridge for hospitality when this work was completed.
We acknowledge support from S.~Rouberol for running the cluster for us.

\bibliography{LSStructure}


\appendix

\section{Radii decimations}
\label{decimation}

The purpose of this Appendix is to make sure that the expression of
$\varphi(\{\lambda\})$ is consistent with variable decimation, i.e.  we want to make sure that
\begin{eqnarray}
\hskip -1cm
\varphi(\{\lambda_{1},\dots,\lambda_{n}\})&=& \nonumber
\\
&&\hskip -2cm \varphi(\{\lambda_{1},\dots,\lambda_{n}, \lambda_{n+1}=0,\dots,\lambda_{n+m}=0\})\,,
\label{decimprop}
\end{eqnarray}
where the left hand side is computed from $n$ cells whereas the right hand side is computed with $n+m$ cells.

In order to prove this property,  let us define a set $\mA$ of $n$ cells and a set $\mB$ of $m$ cells.
One can then define the covariance matrix  $\sigma_{ij}(\rho_{i},\rho_{j})$ as in (\ref{sigmaij}) between two any cells of the union of $\mA$ and $\mB$.

We first need to establish a preliminary relation between the element of the inverse matrix $\Xi_{ij}(\{\rho_{k}\})$ and the covariance matrix. 
From
\begin{equation}
\sum_{l=1}^{n}\sigma_{il}(\rho_{i},\rho_{l})\ \Xi_{lj}(\{\rho_{k}\})=\delta_{ij}\,,
\end{equation}
we indeed can derive the following relation,
\begin{equation}
\begin{split}
\hskip -1cm
\sigma_{il}(\rho_{i},\rho_{l})\ \frac{\partial}{\partial \rho_{k}}\left[ \Xi_{lj}(\{\rho_{k}\})\right] \sigma_{jm}(\rho_{j},\rho_{m})
\\
+\frac{\partial}{\partial \rho_{k}}\left[\sigma_{im}(\rho_{i},\rho_{m})\right]=0\,,
\end{split}
\label{VarRelationM}
\end{equation}
where all the repeated indices run from 1 to $n+m$.
One can also write this relation when the inverse matrix is defined from the covariance matrix of the cells restricted in $\mA$
only. Let us define by $\hXi_{\mu\nu}(\{\rho_{\rho \mu}\})$ this matrix and in the following restrict the greek indices  from $1$
to $n$. The previous relation is then transformed into
\begin{equation}
\begin{split}
\sigma_{\mu\lambda}(\rho_{\mu},\rho_{\lambda})\ \frac{\partial}{\partial \rho_{\kappa}}\left[ \hXi_{\lambda\nu}(\{\rho_{\mu}\})\right] \sigma_{\nu\sigma}(\rho_{\nu},\rho_{\sigma})+\\
\frac{\partial}{\partial \rho_{\kappa}}\left[\sigma_{\mu\sigma}(\rho_{\mu},\rho_{\sigma})\right]=0.
\end{split}
\label{VarRelationN}
\end{equation}
The cumulant generating functions for the $n$ cells in $\mA$ is given by
\begin{equation}
\hvarphi(\{\lambda_{\mu}\})=\lambda_{\mu}\tau_{\mu}-\frac{1}{2}\hXi_{\mu\nu}\tau_{\mu}\tau_{\nu}\,,
\end{equation}
with the stationary conditions 
\begin{equation}
\lambda_{\kappa}=\hXi_{\mu\kappa}\tau_{\mu}\frac{\dd \tau_{\kappa}}{\dd \rho_{\kappa}}+
\frac{1}{2}\frac{\partial\hXi_{\mu\nu}}{\partial\rho_{\kappa}}\tau_{\mu}\tau_{\nu}.
\label{lambdakappaexp}
\end{equation}
The purpose of the following calculation is to show that it is identical to the expression of $\varphi({\lambda_{i}})$ describing the cumulant 
generating function of the $n+m$ cells when the last $m-n$ values of $\lambda_{i}$ are set to zero. In this case we have
\begin{equation}
\varphi(\{\lambda_{\mu},0\})=\lambda_{\mu}\tau_{\mu}-\frac{1}{2}\Xi_{ij}\tau_{i}\tau_{j}\,,
\end{equation}
with the stationary conditions 
\begin{eqnarray}
\lambda_{\kappa}&=&\Xi_{i\kappa}\tau_{i}\frac{\dd \tau_{\kappa}}{\dd \rho_{\kappa}}+
\frac{1}{2}\frac{\partial\Xi_{ij}}{\partial\rho_{\kappa}}\tau_{i}\tau_{j};\label{statAll1}\\
0&=&\Xi_{k i}\tau_{i}\frac{\dd \tau_{k}}{\dd \rho_{k}}+
\frac{1}{2}\frac{\partial\Xi_{ij}}{\partial\rho_{k}}\tau_{i}\tau_{j}\,,\label{statAll2}
\end{eqnarray}
for $k$ running from $n+1$ to $n+m$. The second set of constraints allows to determine the values of $\tau_{i}$ for $i\in[n+1,n+m]$
in terms of $\tau_{\nu}$.  It is given by 
\begin{equation}
\htau_{i}=\sigma_{i\mu}\hXi_{\mu\nu}\tau_{\nu}\,,
\end{equation}
where once again repeated greek indices are summed over from $1$ to $n$. This expression is actually valid for any values
of $i$ as when $i$ is in the $1$ to $n$ range we identically have $\htau_{i}=\tau_{i}$.
One can indeed check that for this expression the two terms in Eq. (\ref{statAll2}) are identically 0: indeed
$\Xi_{ki}\htau_{i}=\delta_{k\mu}=0$ for $k\in[n+1,n+m]$ and 
${\partial\Xi_{ij}}/{\partial\rho_{k}}\,\htau_{i}\htau_{j}=
{\partial\Xi_{ij}}/{\partial\rho_{k}}\,\sigma_{i\mu}\hXi_{\mu\nu}\tau_{\nu}
\sigma_{j\mu'}\hXi_{\mu'\nu'}\tau_{\nu'}=
-{\partial\sigma_{\mu\mu'}}/{\partial\rho_{k}}\,\hXi_{\mu\nu}\tau_{\nu}
\hXi_{\mu'\nu'}\tau_{\nu'}=0$ for $k\in[n+1,n+m]$.
Then replacing using this expression for the $\tau_{i}$ in Eq.~(\ref{statAll1}) one gets
\begin{equation}
\lambda_{\kappa}=\Xi_{i\kappa}\sigma_{i\mu}\hXi_{\mu\nu}\tau_{\nu}\frac{\dd \tau_{\kappa}}{\dd \rho_{\kappa}}+
\frac{1}{2}\frac{\partial\Xi_{ij}}{\partial\rho_{\kappa}}\sigma_{i\mu}\hXi_{\mu\nu}\tau_{\nu}
\sigma_{j\mu'}\hXi_{\mu'\nu'}\tau_{\nu'}\,. \nonumber
\end{equation}
Its first term can be simplified using the definition of $\hXi$ and the second by the subsequent use of 
Eqs. (\ref{VarRelationM}) and (\ref{VarRelationN}),
\begin{eqnarray}
\frac{\partial\Xi_{ij}}{\partial\rho_{\kappa}}\sigma_{i\mu}\hXi_{\mu\nu}
\sigma_{j\mu'}\hXi_{\mu'\nu'}
&=&-\frac{\partial}{\partial\rho_{\kappa}}\sigma_{\mu\mu'}\hXi_{\mu\nu}\hXi_{\mu'\nu'}\nonumber\,,\\
&=&\frac{\partial\hXi_{\kappa\sigma}}{\partial\rho_{\kappa}}\sigma_{\kappa\mu}\hXi_{\mu\nu}
\sigma_{\sigma\mu'}\hXi_{\mu'\nu'}\nonumber\,,\\
&=&\frac{\partial\hXi_{\nu\nu'}}{\partial\rho_{\kappa}}\,,
\end{eqnarray}
so that the expression of $\lambda_{\kappa}$ coincides with the expression (\ref{lambdakappaexp}). Finally $\htau_{\mu}=\tau_{\mu}$
ensures that the property (\ref{decimprop}) is valid.

\section{Integration in the complex plane}
\label{Integrationinthecomplexplane}

\subsection{Numerical algorithm}

\begin{figure}
\centerline{ \psfig{file= 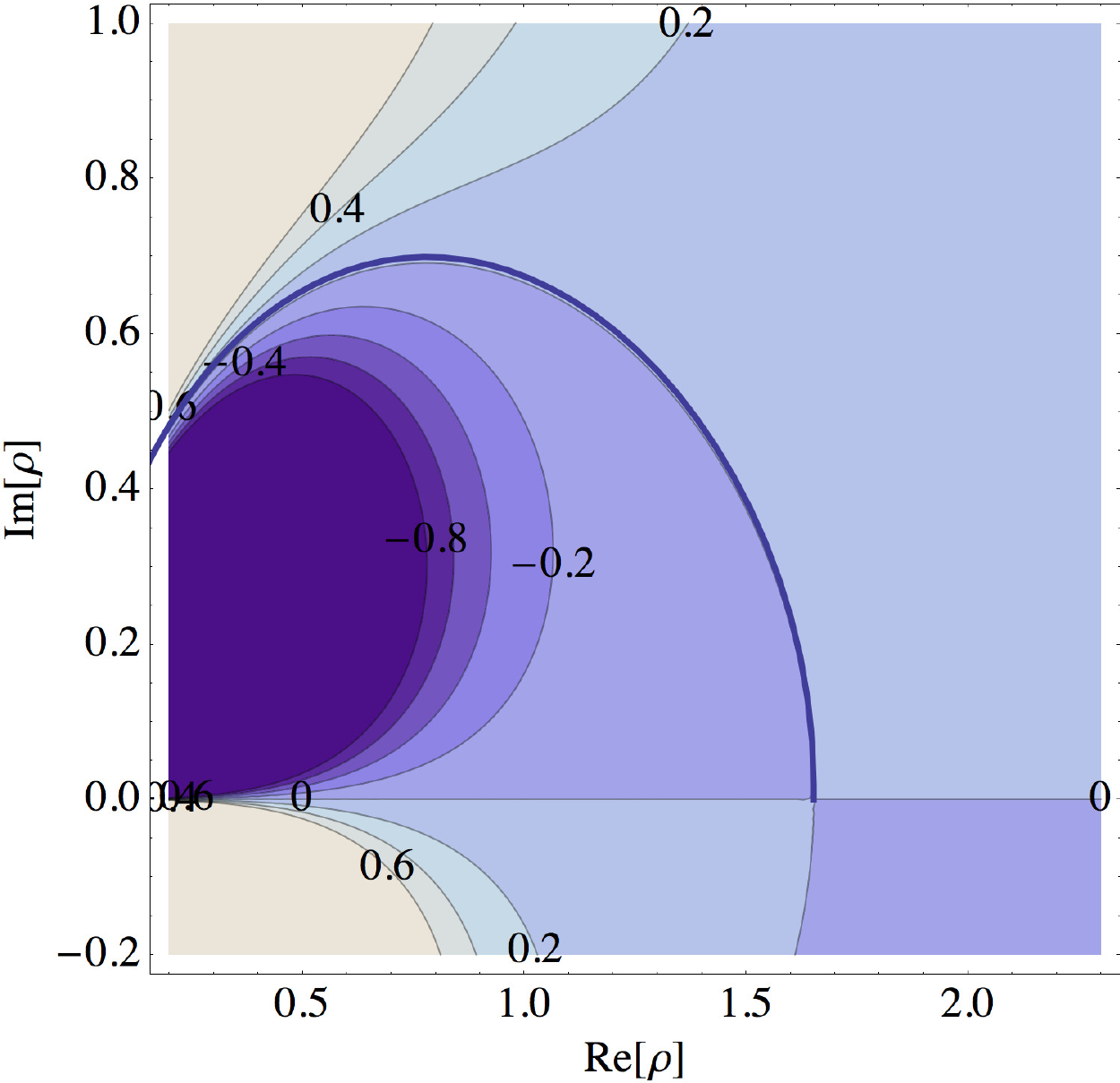, width=8 cm} }
   \caption{The path line in the $\rho$ complex plane. We superimposed the contour plot of the imaginary part of $\varphi(\lambda)-\lambda\hrho$
   to check that it follows a $\Im[\varphi(\lambda)-\lambda\hrho]=0$ line. The starting point on the real axis correspond to the saddle 
   point value.}
   \label{PathLine}
\end{figure}

The computation of the one-point PDF relies on the following expression
\begin{equation}
P(\hrho)=\int_{-\ii\infty}^{+\ii\infty}\frac{\dd\lambda}{2\pi \ii}\exp(-\lambda\hrho+\varphi(\lambda))\,,
\label{InvLapTrans1Db}
\end{equation}
where we explicitly denote $\hrho$ the value of the density for which we want to compute the PDF. This is to distinguish it
from the variable $\rho$ that enters in the calculation of $\varphi(\lambda)$ out of the Legendre transform of $\Psi(\rho)$.
The idea  to achieve   fast convergence of the integral  is to follow a path in the complex plane where the argument of the exponential in 
Eq.~(\ref{InvLapTrans1Db}) is real. The starting point of the calculation is $\rho=\rho_{s}$. When $\hrho$ is small enough (in the regular region)
then we simply have $\rho_{s}=\hrho$ otherwise one should take $\rho_{s}=\rho_{c}$. At this very location, 2 lines of vanishing imaginary part
of $-\lambda\hrho+\varphi(\lambda)$ cross, one along the real axis (obviously) and one parallel to the imaginary axis (precisely because we are at
a saddle point position). The idea is then to build, step by step, 
a path by imposing
\begin{equation}
\delta\left[\varphi(\lambda)-\lambda\hrho\right] \in \mathbb{R}.
\end{equation}
This condition can be written as an infinitesimal variation of $\lambda$. Recalling that $\dd\varphi(\lambda)/\dd\lambda=\rho(\lambda)$,
for each step we have to impose
\begin{equation}
(\rho-\hrho)\delta\lambda\in \mathbb{R},
\end{equation}
which in turns can be obtained by imposing that the complex argument of $(\delta\rho)$ is that of $\left[(\rho-\hrho)\dd^{2}\Psi/\dd\rho^{2}\right]^{*}$
This
is what we implement in practice. Accurate prediction for the PDFs are obtained with about 50 points along the path line.

\subsection{The large density tails}
\label{tails}

\begin{figure}
\centerline{ \psfig{file= 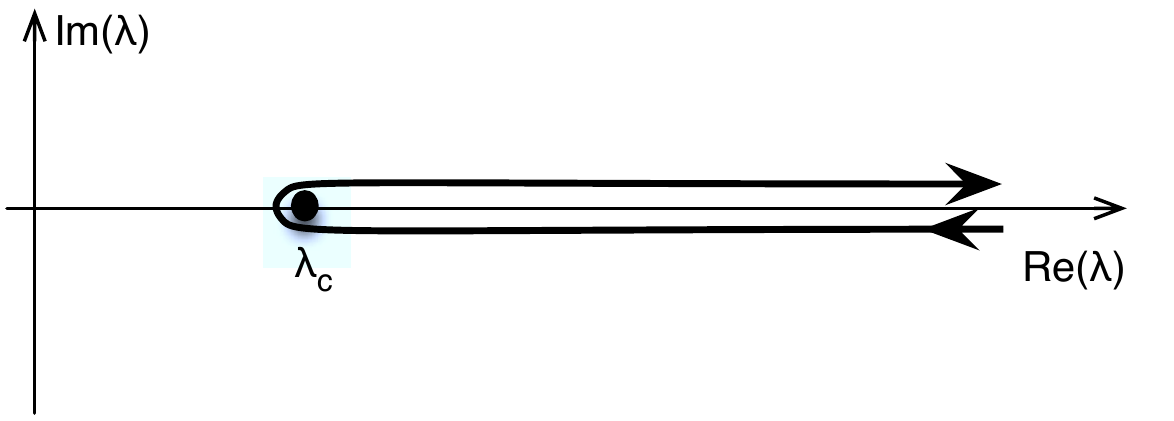, width=6.5 cm} }
   \caption{The path line in the $\lambda$ complex plane  for the computation of the large density asymptotic forms.}
   \label{ContourAsymp}
\end{figure}

The derivation of the rare event tail of the density PDF for large positive densities is based on the
inverse Laplace transform of the generating function $\varphi(\lambda)$ when it is dominated by its
singular part, i.e. for $\lambda\approx \lambda_{c}$. In this case the complex plane contour is pushed 
along the real axis wrapping around the singular value $\lambda_{c}$ as depicted on Fig. 
\ref{ContourAsymp}.

The general form for the density PDF given by Eq. (\ref{InvLapTrans1Db})
is expressed  using the form (\ref{sphiexp}) following the path shown on Fig. \ref{ContourAsymp}. 
As the contributions from the two branches of the path lines are complex conjugate, it eventually leads
to the form,
\begin{eqnarray}
P(\hrho)&\approx&
\Im\left\{
\int_{\ii\epsilon+\lambda_{c}}^{\ii\epsilon+\infty}
\frac{\dd\lambda}{\pi}
\exp[{\varphi_{c}-\lambda_{c}\hrho-(\lambda-\lambda_{c})(\hrho-\rho_{c})}]\right.\nonumber\\
&&
\times \left.\left[1+a_{3/2}(\lambda-\lambda_{c})^{3/2}+\dots
\right]\right\}\,,
\end{eqnarray}
where we keep only the dominant singular part in $\varphi(\lambda)$ and where $\Im$ denotes the imaginary part. This integral can easily
be computed and it leads to,
\begin{equation}
P(\hrho)\approx\exp\left(\varphi _{c}- \lambda _{c}\hrho \right)
\left(\frac{3\, \Im{(a_{\frac{3}{2}} )} }{4 \sqrt{\pi } \left(\hrho -\rho _{c}\right)^{5/2}}+\dots\right).
\end{equation}
Subleading contributions can be computed in a similar way when $\exp(\varphi(\lambda))$ is expanded to higher
order.  Note that by symmetry, only half integer terms that appear in this expansion will actually 
contribute.
\section{The constrained variance}
\label{Ap:ConsVar}

In this appendix we complement the calculations started in Subsect.~\ref{sec:ConsProf} where we computed
the expected slope under a local density constraint. Pursuing along the same line of calculations, 
the variance of $\hrho_{2}$ given $\hrho_{1}$  can be computed from the conditional value of $\hrho_{2}^{2}$. 
It is given by the second derivative of the moment generating function, and is therefore given by
\begin{equation}
\begin{split}
\int\dd \hrho_{2} \ \hrho_{2}^{2} \ P(\hrho_{1},\hrho_{2})= \hskip 4cm\\
\int_{-\ii\infty}^{+\ii\infty}\frac{\dd\lambda_{1}}{2\pi \ii}
\left[\left.\frac{\partial^{2} \varphi(\lambda_{1},\lambda_{2})}{\partial\lambda_{2}^{2}}\right\vert_{\lambda_{2}=0}
\right. \nonumber \\
\left.+\left(\left.\frac{\partial \varphi(\lambda_{1},\lambda_{2})}{\partial\lambda_{2}}\right\vert_{\lambda_{2}=0}
\right)^{2}\right]
\exp(-\lambda_{1}\hrho_{1}+\varphi(\lambda_{1}))\,.
\end{split}
\end{equation}
The calculation of its approximate form in the low-$\rho$ saddle point limit is a bit more cumbersome. 
Indeed, in the low
variance limit in which this approximation is derived the two terms in the square brackets are not of the same order, the first being subdominant with respect the second. It is nonetheless possible to compute the resulting \emph{cumulant} in the low density limit. 
Formally, differentiating Eq.~(\ref{eq1storder}) with respect to $\lambda_2$ we have
\begin{equation}
\frac{\partial^{2} \varphi(\lambda_{1},\lambda_{2})}{\partial\lambda_{2}^{2}}=
\frac{\partial \rho_{2}(\lambda_{1},\lambda_{2})}{\partial\lambda_{2}}\,,
\end{equation}
 from the Legendre stationary condition,
which, after inversion of the partial derivatives, is formally given by
\begin{equation}
\frac{\partial^{2} \varphi(\lambda_{1},\lambda_{2})}{\partial\lambda_{2}^{2}}=
\frac{\Psi_{,\rho_{1}\rho_{1}}}{\Psi_{,\rho_{1}\rho_{1}}\Psi_{,\rho_{2}\rho_{2}}-\Psi^{\ \ 2}_{,\rho_{1}\rho_{2}}}\,,
\end{equation}
where $\Psi_{,\rho_{i}\rho_{j}}\equiv \partial^{2}\Psi/\partial\rho_{i}\partial\rho_{j}$ are calculated 
at the stationary point.
On the other hand 
${\partial\varphi(\lambda_{1},\lambda_{2})}/{\partial\lambda_{2}}$ can be expanded as 
\begin{equation}
\begin{split}
\frac{\partial\varphi(\lambda_{1},\lambda_{2})}{\partial\lambda_{2}}
= 
\varphi(\lambda_{s},0)+(\lambda_{1}-\lambda_{s}) \times \hskip 2 cm\\
\frac{\partial^{2}\varphi(\lambda_{1},\lambda_{2})}{\partial\lambda_{2}\partial\lambda_{1}}
+\frac{1}{2}(\lambda_{1}-\lambda_{s})^{2}
\frac{\partial^{3}\varphi(\lambda_{1},\lambda_{2})}{\partial\lambda_{2}\partial\lambda_{1}^{2}}
+\dots
\end{split}
\end{equation}
near the saddle point value $\lambda_{s}$. The integration of ${\partial\varphi(\lambda_{1},\lambda_{2})}/{\partial\lambda_{2}}$ 
in the complex plane therefore leads to a correction from the $(\lambda_{1}-\lambda_{s})^{2}$ term. 
It can be  verified though that this 
contribution vanishes when one takes the cumulant. The integration of 
$\left({\partial\varphi(\lambda_{1},\lambda_{2})}/{\partial\lambda_{2}}\right)$ however leads to an extra term
due to the second term in the previous expansion. The resulting term reads $
\left[{\partial^{2}\varphi}/{\partial\lambda_{2}\partial\lambda_{1}}\right]^{2}/
{\partial^{2}\varphi}/{\partial\lambda_{1}^{2}}$,
so that
\begin{equation}
\begin{split}
\langle\rho_{2}^{2}\rangle_{\rho_{1}}-\langle\rho_{2}\rangle^{2}_{\rho_{1}}=
\frac{\partial^{2}\varphi}{\partial\lambda_{1}^{2}}-
\left[\frac{\partial^{2}\varphi}{\partial\lambda_{2}\partial\lambda_{1}}\right]^{2}/
\frac{\partial^{2}\varphi}{\partial\lambda_{1}^{2}}
=\\
\left[
\frac{\partial^{2}\varphi}{\partial\lambda_{1}^{2}}
\frac{\partial^{2}\varphi}{\partial\lambda_{2}^{2}}
-
\left(\frac{\partial^{2}\varphi}{\partial\lambda_{1}\partial\lambda_{2}}\right)^{2}
\right]/\frac{\partial^{2}\varphi}{\partial\lambda_{1}^{2}}\,,
\end{split}
\end{equation}
which can be rewritten more compactly as
\begin{equation}
\langle\hrho_{2}^{2}\rangle_{\hrho_{1}}-\langle\hrho_{2}\rangle^{2}_{\hrho_{1}}=
{1}/{\left.\Psi_{,\rho_{2}\rho_{2}}\right\vert_{\hrho_{1},\overline\rho_{2}(\hrho_{1})}}\,,
\label{saddlevarprof}
\end{equation}
when expressed in terms of $\Psi$.

\section{Simulations}
\label{Simulations}
For the purpose of this paper, we have carried out a dark matter simulation with {\tt Gadget2} \citep{gadget2}.
This simulation is characterized by the following $\Lambda$CDM cosmology: $\Omega_{\rm m}=0.265 $, $\Omega_{\Lambda}=0.735$, $n=0.958$, $H_0=70 $ km$\cdot s^{-1} \cdot $Mpc$^{-1}$ and $\sigma _8=0.8$, 
$\Omega_{b}=0.045$
within one standard deviation of WMAP7 results \citep{wmap7}. 
The box size is 500 Mpc$/h$ sampled with  $1024^3$ particles, the softening length 24 kpc$/h$.
Initial conditions are generated using {\tt mpgrafic}  \citep{mpgrafic}.
The variances and running indexes are measured from the  theoretical  power spectra produced by  {\tt mpgrafic}. 
Snapshots are saved for $z=0, 0.65, 0.97, 1.46, 2.33$ and $3.9$.
An Octree is built for each snapshot, which allows us to count very efficiently all particles 
within a given  sequence of concentric spheres of radii between $R=4,
5 \cdots $ up to $ 18 {\rm Mpc}/h $.  The center of these spheres is sampled regularly on a grid of 
$ 10\, {\rm Mpc}/h $ aside, leading to 117649 estimates of the density per snapshot. All histograms drawn in this paper 
are derived from these samples.  Note that the cells overlap for radii larger than $10 \, {\rm Mpc}/h $.

\section{Systematic comparisons with simulation}

We collect here figures that are too large to be put in the main text.

\begin{figure*}
\centerline{ \psfig{file=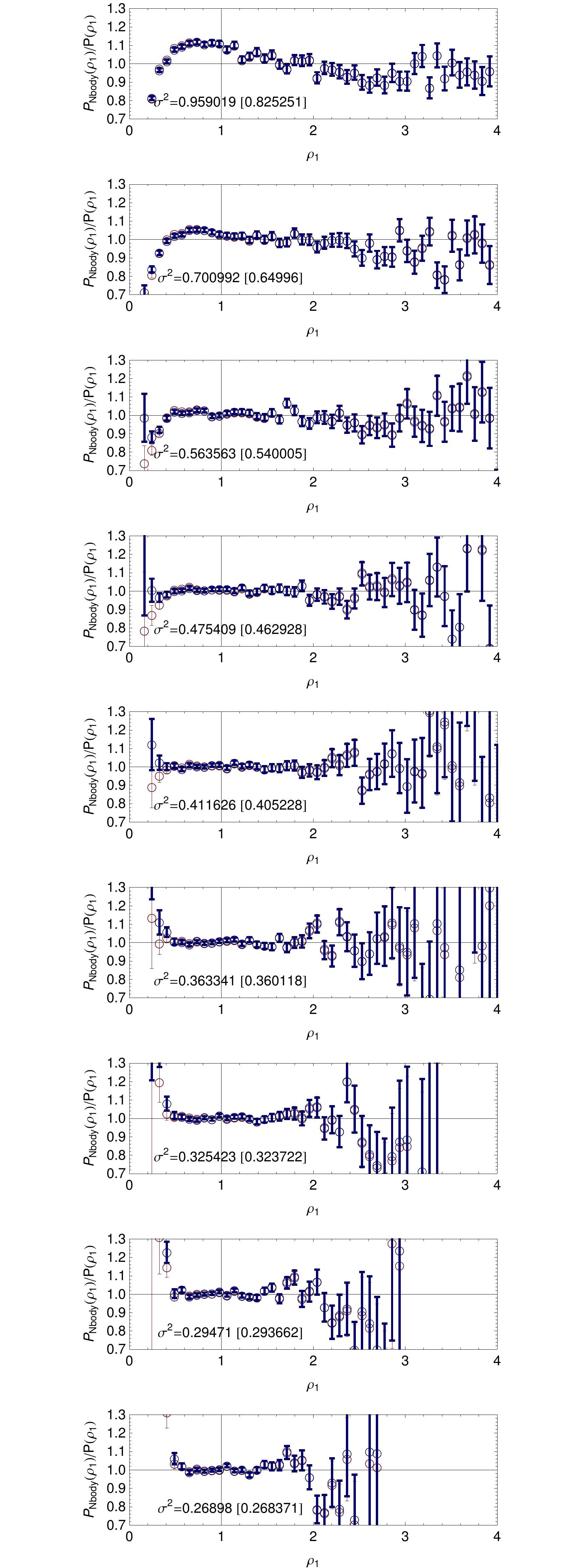, width=7 cm}   \psfig{file=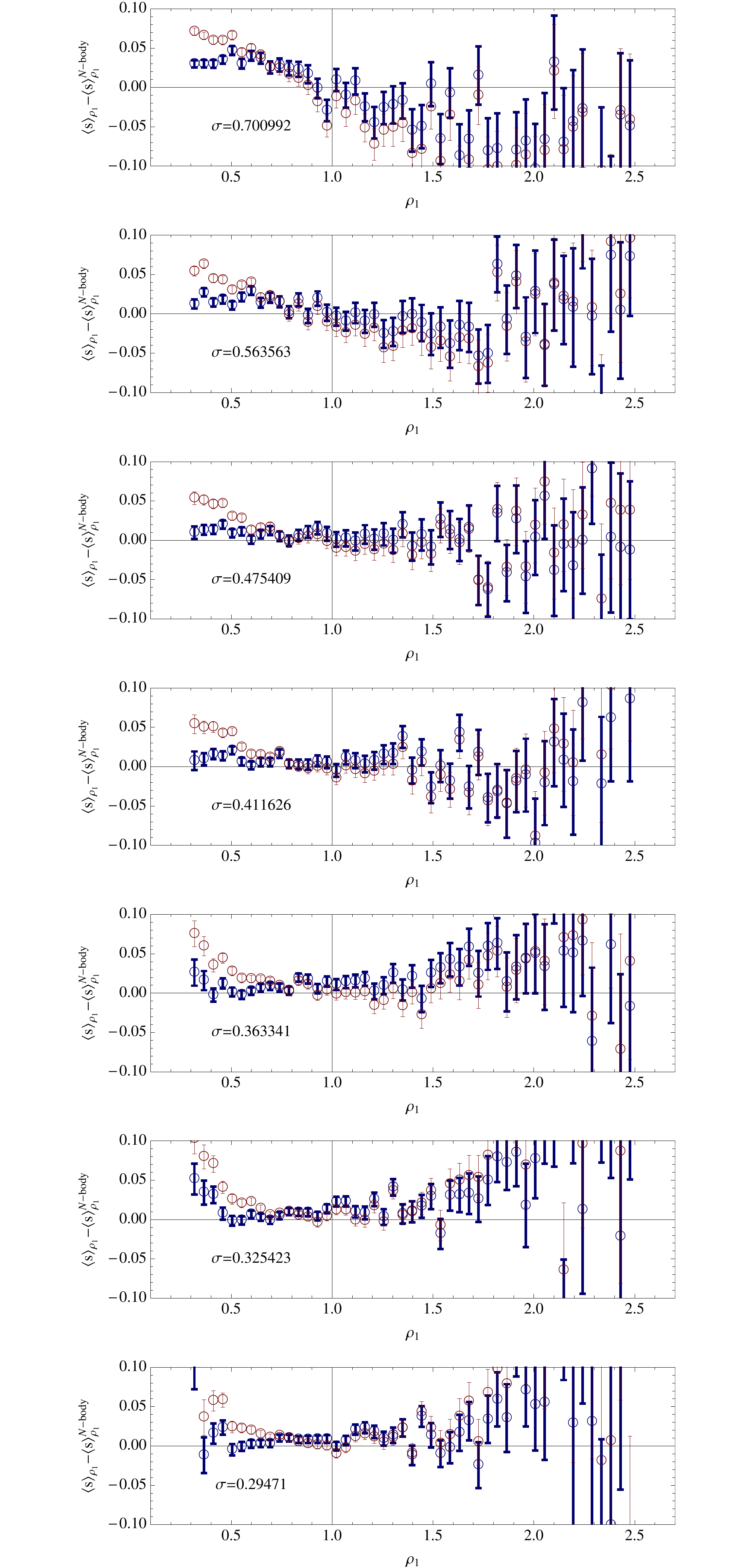, width=7 cm}}
   \caption{Left panel: the residuals of the expected density PDF from smoothing scale of  $R_{1}=4$ (top) to $R_{1}=20\,h^{-1}$Mpc. This is for $z=0.97$ (same convention as in Fig. \ref{ExpPDF10}).
   The values of $\sigma^{2}$ at the smoothing scale is given in the inset. The value in square bracket is the linear value. Right panel: the residuals for the expected average profile between scale $R_{1}$ and $R_{2}=R_{1}+1 h^{-1}$Mpc. 
   From top to bottom we have $R_{1}=6$ to $18\,h^{-1}$Mpc. Same convention as in Fig. \ref{PDFSlope}.}
   \label{ExpPDF-zp97-rest}
\end{figure*}
\vfill
\end{document}